\documentclass[pre,superscriptaddress,twocolumn]{revtex4-1}
\pdfoutput=1
\usepackage[colorlinks=true,urlcolor=blue]{hyperref}
\usepackage{amsfonts}
\usepackage{graphicx}
\usepackage{mathrsfs}
\usepackage{amsmath}
\usepackage{xcolor}
\usepackage{bm}

\usepackage{color}
\definecolor{red}{rgb}{0.75,0,0}
\definecolor{blue}{rgb}{0,0,0.75}
\definecolor{green}{rgb}{0,0.5,0}

\newcommand{\red}[1]{{\color{red} #1}}

\begin{document}

\title{Cross-talk between topological defects in different fields \\ revealed by nematic microfluidics}

\author{Luca Giomi}
\affiliation{Instituut-Lorentz, Universiteit Leiden, P.O. Box 9506, 2300 RA Leiden, The Netherlands}
\author{\v{Z}iga Kos}
\affiliation{University of Ljubljana, Faculty of Mathematics and Physics, Jadranska 19, 1000 Ljubljana, Slovenia}
\author{Miha Ravnik}
\affiliation{University of Ljubljana, Faculty of Mathematics and Physics, Jadranska 19, 1000 Ljubljana, Slovenia}
\affiliation{Institut Jo\v{z}ef Stefan, Jamova cesta 39, 1000 Ljubljana, Slovenia}
\author{Anupam Sengupta}
\email{anupams@ethz.ch}
\affiliation{Institute for Environmental Engineering, Department of Civil, Environmental and Geomatic Engineering, ETH Zurich, Stefano-Franscini-Platz 5, 8093 Zurich, Switzerland}
\affiliation{Max Planck Institute for Dynamics and Self-Organization (MPIDS), Am Fassberg 17, 37077 G{\"o}ttingen, Germany}

\begin{abstract}
Topological defects are singularities in material fields that play a vital role across a range of systems: from cosmic microwave background polarization to superconductors, and biological materials. Although topological defects and their mutual interactions have been extensively studied, little is known about the interplay between defects in different fields -- especially when they co-evolve -- within the same physical system. Here, using nematic microfluidics, we study the cross-talk of topological defects in two different material fields -- the velocity field and the molecular orientational field. Specifically, we generate hydrodynamic stagnation points of different topological charges at the center of star-shaped microfluidic junctions, which then interact with emergent topological defects in the orientational field of the nematic director. We combine experiments, and analytical and numerical calculations to demonstrate that a hydrodynamic singularity of given topological charge can nucleate a nematic defect of equal topological charge, and corroborate this by creating $-1$, $-2$ and $-3$ topological defects in $4-$, $6-$, and $8-$arm junctions. Our work is an attempt toward understanding materials that are governed by distinctly multi-field topology, where disparate topology-carrying fields are coupled, and concertedly determine the material properties and response.
\end{abstract}

\maketitle

\section{Introduction}

Defects are ubiquitous in nature and lie at the heart of numerous physical mechanisms, including, melting in two-dimensional crystals~\cite{Nelson:1979} to cosmic strings and other topological defects in the early universe~\cite{Kibble:1976, Ade:2013}. Vortices are possibly the most common examples of defects in flowing media \cite{Newton:1713,Batchelor:1967}. In a typical hydrodynamic vortex, the fluid velocity, $\bm{v}$, rotates by $2\pi$ along any closed loop around the vortex core, and has an undefined direction at the core. More generally, topological defects are singular points or lines in a distinct scalar, vector or tensor field that can be characterized by topological invariants, including winding number (or index) for two-dimensional, and topological charge for three-dimensional variations of the fields~\cite{Alexander:2012, Mermin:1979}. Topological defects have been long known to mediate key processes in a wide range of settings, including knotted flow field stream lines~\cite{Kleckner:2013}, defects in light fields~\cite{Desyatnikov:2012}, knotted defect lines in complex fluids~\cite{Tkalec:2011}, defects in Type-2 superconductors~\cite{Tinkham:1997}, spontaneous flow in active fluids \cite{Sanchez:2012,Giomi:2014,Keber:2014,Giomi:2015}, and even in conduction properties of electron nematics~\cite{Carlson:2010}.

%efects are ubiquitous in nature and lie at the heart of several physical mechanisms, from melting in two-dimensional crystals \cite{Nelson:1979} to symmetry breaking in the early universe \cite{Kibble:1976}. Vortices are possibly the most common example of defects in flowing media and, for centuries, they have stimulated the imagination of scientists, mathematicians and philosophers, as exemplified by the debate between Newton and Descartes on the existence of ``swirling vortices'' guiding motion of celestial bodies in the solar system \cite{Newton:1713}. In a typical hydrodynamic vortex, the fluid velocity $\bm{v}$ rotates by $2\pi$ along any closed loop surrounding the vortex core and has an undefined orientation at the core. More generally, dynamical defects are singular points or lines in a flow around which the velocity field rotates by $2\pi k$, where $k$ is an integer commonly referred to as index, winding number or topological charge. These can be either vortices ($k=1$), multiplets of vortices ($k=2,\,3\ldots$) or saddle stagnation points of various degree of rotational symmetry ($k=-1,\,-2\ldots$) \cite{Batchelor:1967}.

The interaction between topological defects is governed by the defect topology and the underlying energetics. %Between defects of same sign, the interaction is repulsive in nature, and, attractive between opposite sign defects. 
Similar to electrically charged particles, like-sign topological defects, in general, repel each other, while defects of opposite sign attract. However, this can be additionally affected by the geometry and surface properties of the environment~\cite{Sharifi-Mood:2015,Wei:2016}, and the presence of an external stimulus~\cite{Chuang:1991, Toth:2002, Giomi:2013, Bowick:2008, Dierking:2012}. 
%conductivity~\cite{Balke:2011}, and rheology~\cite{Wood:2011}.The interaction, governed by the defect topology and the underlying energetics, %affects microscopic as well as the macroscopic attributes of a system~\cite{Stratford:2014, Liew, Balke:2011, Wood:2011}.The interaction is repulsive between defects of same sign and attractive between opposite sign defects. Furthermore, defect topology and interactions are affected also by the geometry and surface properties of the environment~\cite{Chuang:1991, Toth:2002, Giomi:2013, Dierking:2012}. 
Emergence of topological defects in a field, and the resulting interactions between them, have been well characterized~\cite{Chaikin:2000}. Yet, how topological defects in a system, can co-evolve in and interact across disparate fields, is largely unexplored. It is rather recent that multi-field topological interactions were demonstrated in optics where singularities in optical birefringence created topological defects in the light field~\cite{Brasselet:2012, Cancula:2014}. The growing evidence that topological defects perform vital biological functions~\cite{Peng:2016,Saw:2017,Kawaguchi:2017}, creates a fundamental need for an integrated understanding of defect interactions, especially in relation to those in a different field, for instance, in the surrounding micro-environment.

\begin{figure*}[t]
\centering
\includegraphics[width=1.0\textwidth]{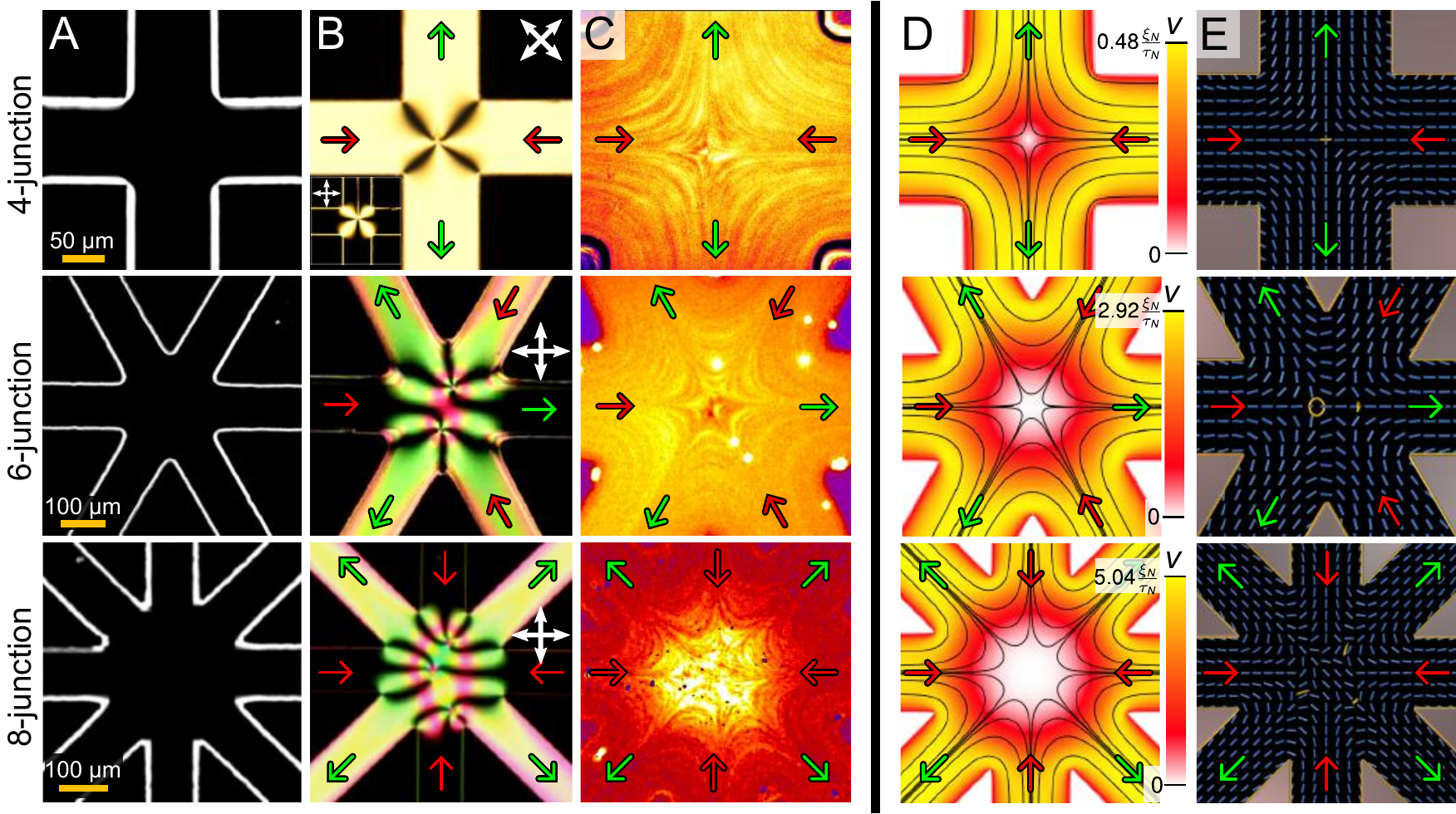}
\caption{\label{fig:1}Emergence of nematic topological defects and hydrodynamic singularities at a microfluidic junction. (A) Generic star-shaped microfluidic junctions. From top: $4-$, $6-$, and $8-$arm microfluidic junctions, and corresponding POM images (B) of the emergent topological defects at the center: $-1$ ($4-$arm junction), $-2$ defect split into two $-1$ defects ($6-$arm junction), and $-3$ defect split into three $-1$ defects ($8-$arm junction). The double headed arrows indicate the orientation of the crossed-polarizers. The inflow and outflow arms are indicated by the red and green arrows respectively. (C) Epi-fluorescent imaging of flowing fluorescent tracers reveal the hydrodynamic stagnation points at the geometric centers of each microfluidic junction. (D and E) Nematic flows at microfluidic junctions reproduced in numerical simulations. (D) Streamlines of flow profile in simulations. The range in which the velocity magnitude is drawn is given in units of nematic correlation length, $\xi_N$, divided by the characteristic nematic time scale, $\tau_N$ (see also Appendix \ref {sec:numerical_simulations}). (E) The director profile (blue rods) and the isosurface of the nematic scalar order parameter, drawn at $S=0.4$ in yellow, as corresponding to the flow field in (D).}	
\end{figure*}

Complex nematic fluids have proven to be a versatile test-bed for studying, testing and realizing diverse topological concepts~\cite{Sec:2014, Martinez:2014, Wang:2016}, owing primarily to their inherent softness and strong response to external stimuli, and in context of the present work, their material fluidity~\cite{Foster:1971, DeGennes:1995}. Liquid crystal microfluidics~\cite{Sengupta:2014} has emerged as a potent toolkit to modulate fluid and material structures due to the coupling between the two main material fields -- the fluid velocity field and the molecular orientational field (director)~\cite{DeGennes:1995}. The flow-director coupling regulates transport properties of nematic suspensions~\cite{Navarro:2014,Stark:2001}, tunes the rheology of the LC fluids~\cite{Henrich:2013, Cordoba:2016, Batista:2015, Sengupta:2013b}, and mediates annihilation-creation dynamics of topological defects~\cite{Thampi:2015, Giomi:2013}. Microfluidics based on complex anisotropic fluids has allowed for potential applications~\cite{Tiribocchi:2014}, and novel designs of micro-cargo transport~\cite{Sengupta:2013}, tunable fluid resistivity~\cite{Na:2010}, color filters~\cite{Cuennet:2013}, and bio-chemical sensors~\cite{Liu:2012}. 

In this paper, we study the emergence of topological defects in two different fields present in the nematic microfluidic system: the stagnation point, a hydrodynamic singularity in the flow velocity field, and the nematic defect, a topological singularity in the molecular orientatation field. We characterize the cross-interaction between these topological defects using star-shaped microfluidic junctions and flowing nematic fluid \red{(Fig. \ref{fig:1}A)}. We show that the nucleation, and the nature of the nematic defects, are determined by the topology of the flow defect, such that a hydrodynamic stagnation point of topological charge $1-N/2$, with $N$ the number of arms of the junction, nucleates a defect of the same topological charge in the nematic director field. The multi-field defect interaction is underpinned by the coupling between the two fields, which we tune via microfluidic geometry, and the nematic flow parameters.  
%topological defect of charge $n$ in the flow field induces nematic defect of equal $n$-charge. 
We observe transformations between topological states, including the decay of nematic defects to lower topological charges. Notably, the reconfiguration time scales for the defects from the two different fields -- $\sim10^{-5}\,\text{s}$ for the hydrodynamic stagnation points, and $\sim1\,\text{s}$ for the nematic defects -- are resolved, and possible ramifications of this separation of time scales are discussed. Finally, this work is a realization of a material system governed by the topology of multiple coupled fields -- a platform which can be extended further, potentially leading to the development of new topological materials or topological material phenomena.

\section{Tuning Topology with Hydrodynamics}

We study the emergence of topological defects using a combination of experiments, numerical modeling and theory. Experimentally, we employ star-shaped microfluidic junctions, fabricated by soft lithography techniques (Materials and Methods, and Appendix \ref{sec:experimental_setup}). In theory and modeling, we use phenomenological Beris-Edwards model type approach based on the nematic order parameter tensor -- a strong tool to study nematic structures, especially defects at mesoscopic scale \cite{DeGennes:1995}. 

The cross-interaction between the velocity and the nematic fields is governed by an interplay of multiple effects: material viscosity, nematic elasticity, channel dimensions, and the strength of the flow (Fig. \ref{fig:1}A). The combined effect is captured by a single dimensionless number, the Ericksen number, $\mathrm{Er} = \eta v l/K$ \cite{Oswald:2005}; $\eta$ being the viscosity, $v$, the flow velocity, $l$, the channel hydraulic diameter, and $K$, the 5CB elastic constant (Appendix \ref{sec:experimental_setup}). The Ericksen number ($0.4 \lesssim \mathrm{Er} \lesssim 70$ in our experiments), thus gives a relative measure of the viscous and elastic stresses.

%\red{The central control parameter in our experiments is the so called Ericksen number \cite{Oswald:2005} expressing the competition between viscous and elastic stresses:}
%Here $0.4 \lesssim \mathrm{Er} \lesssim 70$ 

%We study the emergence of the topological defects at the microfludic junction over a range of Ericksen numbers, $\mathrm{Er} = \eta v l/K$, of the nematic flow in the channel arms \cite{Oswald:2005}. Here, $\eta$ and $K$ are the dynamic viscosity and elastic constant of nematic 5CB, $v$ is flow velocity, and $l$ the hydraulic diameter of microchannel arm (Materials and Methods). 

Fig. \ref{fig:1}B shows the nematic defects obtained in a $4-$, $6-$ and $8-$arm microfluidic junction. In each case, no defect was observed for $\mathrm{Er} < 1$ -- a nematofluidic regime in which the elastic torque far outweighs the viscous torque. In the $4-$arm junction, the first appearance of a $-1$ defect is observed at $\mathrm{Er} = 2$, and is found to stabilize at $\mathrm{Er} > 5$. Fig. \ref{fig:1}B (top panel, imaged at $\mathrm{Er}~\approx 10$) shows polarization optical micrograph (POM) of a stable defect of strength $-1$, at the center of the junction. Increasing the number of arms (Fig. \ref{fig:1}B, middle and lower panels) results in increase in the net topological charge at the junction center: $-2$ (imaged at $\mathrm{Er}~\approx 18$ in $6-$arm channel) and $-3$ (imaged at $\mathrm{Er}~\approx 22$ in $8-$arm channel). High charge defects (greater than $-1$), decayed into multiples of the $-1$ defects: the $-2$ defect into a pair of $-1$ defects, and the $-3$ defect decayed into three $-1$ defects. By overlaying the positions of the hydrodynamic and nematic topological defects, we find that in a $4-$arm junction, the $-1$ defect lies within a micrometer from stagnation point. When averaged over time, the positions of the topological defects coincided. Similarly, in the $6-$ and the $8-$arm junctions, the defects of higher charge (existing as multiples of $-1$ monopole), are found to fall within a stagnation zone, a region at the junction center where the speed was less than $10\%$ of the far field value. 

We have reproduced the experimental results {\em in silico} using numerical simulations of three-dimensional microfluidic junctions based the Navier-Stokes equation coupled with the Beris-Edwards equations of nematodynamics \cite{Beris:1994} (see Materials \& Methods, and Appendix \ref{sec:numerical_simulations}). Figs. \ref{fig:1}D and E show the numerical flow velocity and the nematic ordering at the $4-$, $6-$ and $8-$arms junctions. The isosurfaces of the nematic order parameter (Fig. \ref{fig:1}E) show stable $-1$ defect loops, i.e. defects of charge $-1$ consisting of a disclination loop with winding number $-1/2$, in good qualitative agreement with the experimental results. Interestingly, the numerical modeling shows that the material flow singularity emerges as a line region, extending from the top to the bottom of the channel, whereas the nematic defects evolve into small loops, which are topologically equivalent to 3D point defects \cite{Wang:2016}. However, from a topological perspective, our setup allows us to fully characterize the 3D nematic defects by an effective 2D invariant, like the winding number. In essence, by simply capturing the mid-plane intersection of the nematic field, we are able to describe the nematic defect, since the channel geometry and the anchoring conditions limit any possible variation of the director field \cite{Pieranski:2016} normal to the mid-plane. To generalize, the demonstrated system gives the cross-interaction between topological line defects and point defects (which at topological level can be considered with 2D invariants), creating an interesting topological test bed with defects of different dimensionality.

\begin{figure}[t]
\centering
\includegraphics[width=\columnwidth]{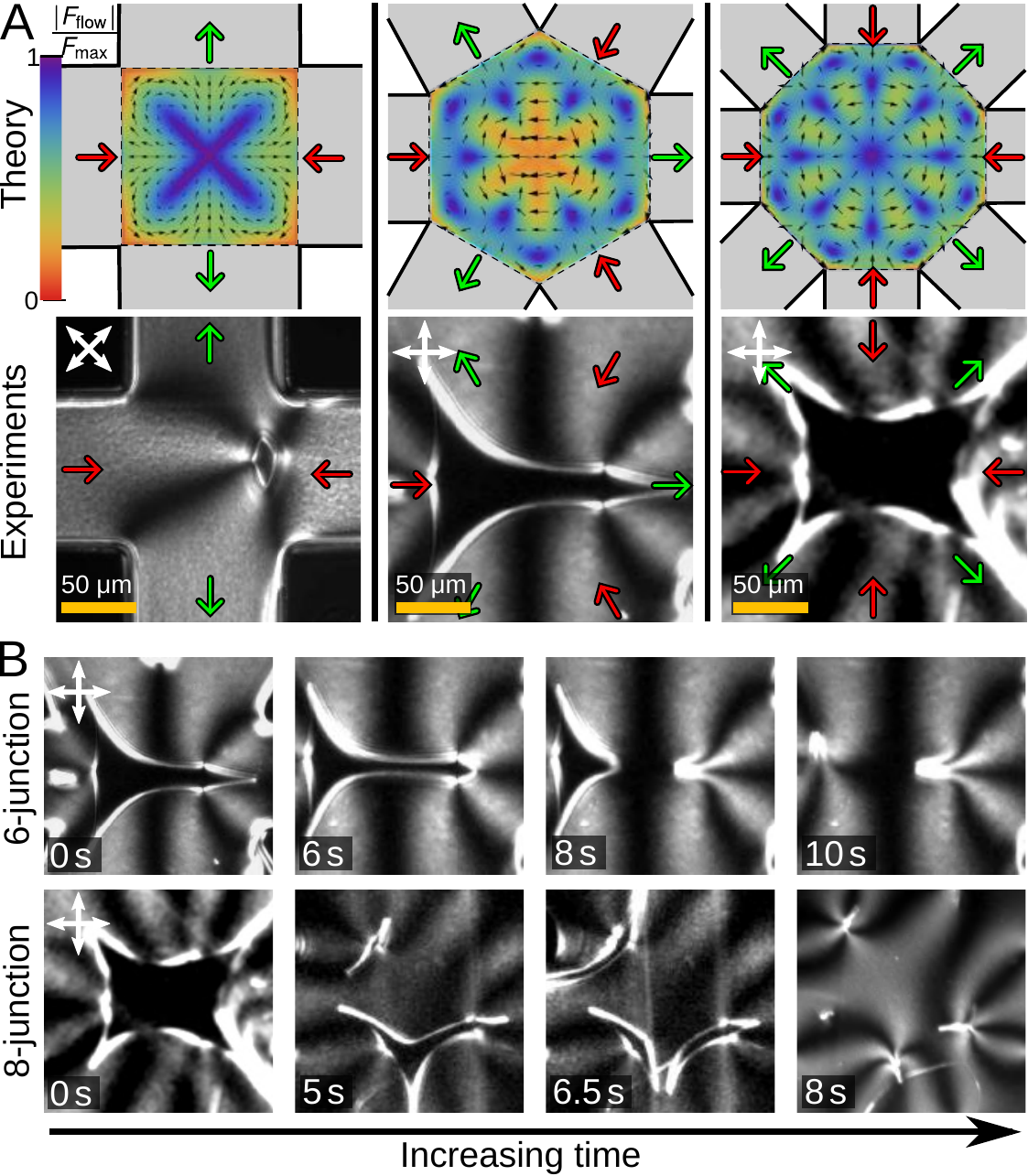}
\caption{\label{fig:2} Fractionalization of topological defects. (A) The hydrodynamic force field experienced by a two-dimensional defect of charge $-1$, $-2$ and $-3$ confined inside a $4-$, $6-$ and $8-$arms junction (upper panel).
%The hydrodynamic force, calculated from Eq. \eqref{eq:force}, results into an effective attraction between the topological defect and the central hydrodynamic stagnation point. 
Lower panel shows POM images of nematic defects right after formation: $-1$ defect loop at a $4-$arm, $-2$ defect at the $6-$arm, and the $-3$ defect at the $8-$arm junctions. In each case, the defect loop encloses a homeotropic domain, close to the junction center, where the surface-induced anchoring remains unperturbed. Outside this domain, the nematic director is aligned due to the flow. (B) The higher strength topological defects decay into multiple defects of charge $-1$, shown here as a time sequence for $-2$ (upper panel) and $-3$ defect (lower panel). The $-1$ loop defect stabilizes by shrinking the enclosed homeotropic domain, and thereby, reducing the effective length of the defect loop (Fig.\ref{fig:3}A).}
\end{figure}

\section{Global Constraints and Local Forces}

The topological structure emerging at the center of the junction, as revealed by our experimental and numerical findings, results from a combination of global topological constraints and local mechanical effects. The shear flow inside the arms tends to align the director along the arms centerline. This drives the formation of $2n$ disclinations of topological charge $+1/2$ at the corners of the $2n-$sided polygon representing the central region of the junction (e.g. top row of Fig. \ref{fig:1}B). The total topological charge of the junction, however, is constrained by the Poincar\'{e}-Hopf theorem \cite{Kamien:2002}, by virtue of which: $\sum_{i} k_{i} = k_{\rm corners} + k_{\rm bulk} = 1$, where the summation runs over all the topological defects in the system. Thus, the topological charge $k_{\rm corners}=n$, introduced by the $2n$ half-strength disclinations located at the corners, must be compensated by a charge $k_{\rm bulk}=1-n$ in the bulk of the junction. In the case of a $4-$arm junction, $k_{\rm corners}=2$ and $k_{\rm bulk}=-1$. For a $6-$arm junction, on the other hand, $k_{\rm corners}=3$ and $k_{\rm bulk}=-2$ and so on. At large Ericksen numbers, this negative topological charge is attracted toward the central stagnation point, due to aligning effect of the flow, at the expense of the system elastic energy. To gain insight on the physical mechanisms behind this process we have looked for defective solutions of the equation governing the dynamics of the nematic director in the presence of a flow \cite{Landau:1986}. For sake of simplicity, we ignore variations in the direction perpendicular to the plane of the junction, so that, the nematic director can be expressed by the two-dimensional vector-field $\bm{n}=(\cos\theta,\sin\theta,0)$. The dynamics of the angle $\theta$ is governed by the following partial differential equation (see Appendix \ref{sec:nematodynamics}):
\begin{equation}\label{eq:theta}
(\partial_{t}+\bm{v}\cdot\nabla)\theta = \frac{K}{\gamma}\nabla^{2}\theta-\omega_{xy}+\lambda(u_{xx}\cos 2\theta-u_{xy}\sin 2\theta),
\end{equation}
where $\bm{v}$ is the flow velocity, $\omega_{ij}=(\partial_{i}v_{j}-\partial_{j}v_{i})/2$ and $u_{ij}=(\partial_{i}v_{j}+\partial_{j}v_{i})/2$ are respectively the vorticity and strain-rate tensor and $\gamma^{-1}$ is the rotational viscosity. The constant $\lambda$ is known as flow-alignment parameter and dictates how the director rotates as effect of a shear flow \cite{DeGennes:1995,Oswald:2005}. For 5CB, $\lambda \approx 1.1$ and the director orients at an angle $\Delta\theta \approx 13^{\circ}$ with respect to the flow \cite{Sengupta:2012b}. 

Now, due to the symmetry of the junction, the flow is approximatively irrotational in proximity of the central stagnation point. In polar coordinates $(r,\phi)$, with $r=0$ representing the center of the junction, an analytical approximation of the flow yields $v_{r}=v_{0}(r/\mathcal{R})^{n-1}\cos n\phi$ and $v_{\phi}=-v_{0}(r/\mathcal{R})^{n-1}\sin n\phi$, with $v_{0}$ the flow speed at the center of the channels and $\mathcal{R}$ a length scale proportional to the channel width (Appendix \ref{sec:flow}). Then, using standard manipulations, one can then prove that, for a perfectly flow-aligning system with $\lambda=1$, the ideal defective configuration $\theta=(1-n)\phi$ is an exact solution of Eq. \eqref{eq:theta}  (Appendix \ref{sec:irrotational}). For $\lambda \gtrsim 1$, the solution departs from this ideal form, but preserves the rotational symmetry.
 
While the existence of a defective equilibrium configuration depends exclusively on the symmetry of the flow in close proximity of the stagnation point, its stability against the elastic forces depends on the structure of the flow over the entire junction. To clarify this point we have introduced an effective particle model for the dynamics of defects in the presence of a generic potential energy field, as that originating from a background flow at sufficiently large Er. Let us consider the generic free energy $\mathscr{F} = \int {\rm d}A\, \left[K|\nabla\theta|^{2}/2+U(\theta)\right]$, where $U(\theta)$ is a potential energy density, possibly due to the interaction with an externally imposed flow, and let us further assume that the system is populated by a given number of topological defects having position $\bm{R}_{i}=(X_{i},Y_{i})$ and topological charge $k_{i}$. Extending a classic approach by Kawasaki \cite{Kawasaki:1984} and Denniston \cite{Denniston:1996}, one can construct an equation of motion for the moving defect in the form (Appendix \ref{sec:defects}):
\begin{multline}\label{eq:eom}
\dot{\bm{R}}_{i}= \bm{v}(\bm{R}_{i})+\mu_{i}\left(2\pi K \sum_{j \ne i}k_{i}k_{j}\,\frac{\bm{R}_{i}-\bm{R}_{j}}{|\bm{R}_{i}-\bm{R}_{j}|^{2}}+\bm{F}_{i}\right),
\end{multline}
were $\mu_{i} \sim 1/(\gamma k_{i}^{2})$ is a mobility coefficient. The second term on the right-hand side of Eq. \eqref{eq:eom}, corresponds to the well known Coulomb-like elastic interaction between topological charges. The third term, on the other hand, is given by $\bm{F}_{i}=-\nabla_{\bm{R}_{i}}\int {\rm d}A\,U(\theta)$, where the integration is performed over a domain punctured at the locations of the defects, and represents the force experienced by a defect moving in a potential energy field. In the presence of hydrodynamic flow, the latter can be calculated in the form  (Appendix \ref{sec:defects}):
\begin{multline}\label{eq:force}
\bm{F}_{i} = k_{i} \int {\rm d}A\,\frac{\bm{\hat{z}}\times (\bm{r}-\bm{R}_{i})}{|\bm{r}-\bm{R}_{i}|^{2}}\\\,\times \left[\omega_{xy}-\lambda(u_{xx}\sin 2\theta-u_{xy}\cos 2\theta)\right],
\end{multline}
where $\theta$ can be approximated as a linear superposition of the local orientations associated with all the defects: i.e. $\theta(\bm{r})\approx\sum_{j}k_{j}\arctan(y-Y_{j})/(x-X_{j})$. Fig. \ref{fig:2}A (upper panel) shows the force field, calculated via Eq. \eqref{eq:force}, experienced by a disclination of topological charge $-1$, $-2$ and $-3$ confined inside a $4-$, $6-$ and $8-$arm junction. The corresponding flow field, thus the tensorial elements $u_{ij}$ and $\omega_{ij}$ in Eq. \eqref{eq:force}, have been analytically approximated based on the rotational symmetry of the junctions and the location of the stagnation points (Appendix \ref{sec:flow}). 

\begin{figure*}[t]
\centering
\includegraphics[width=\textwidth]{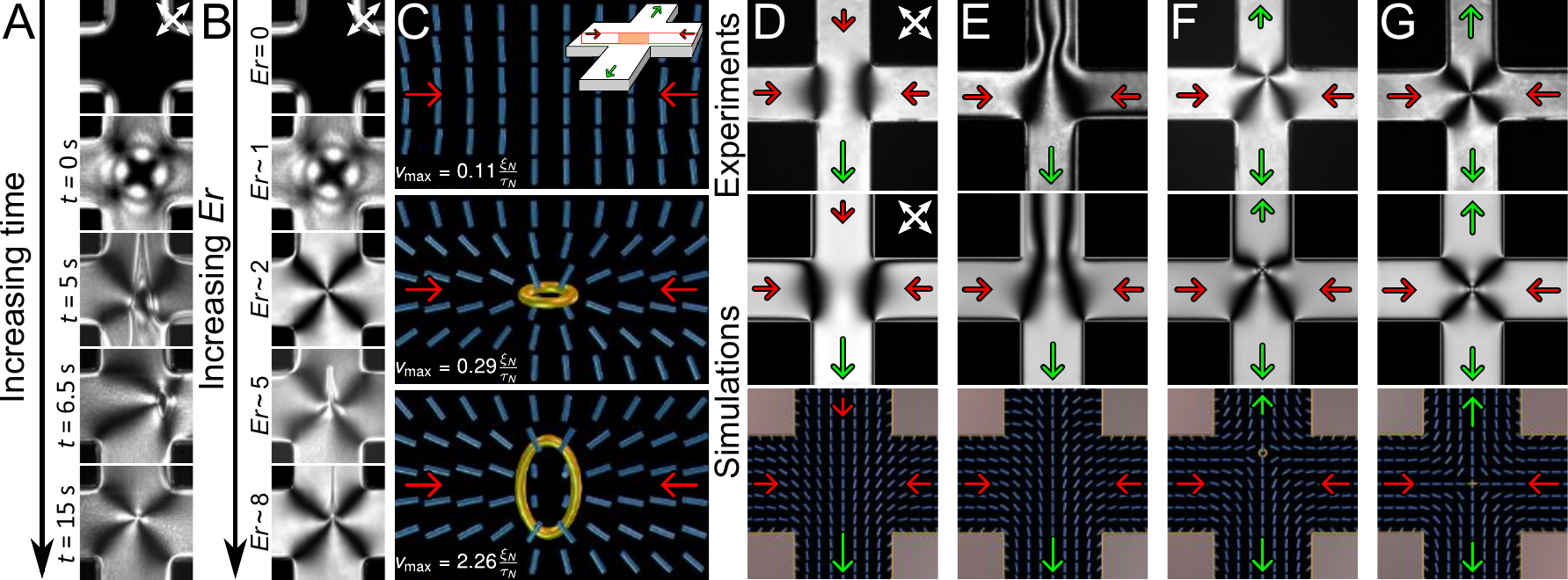}
\caption{\label{fig:3}Dynamics of defect nucleation in a $4-$arm junction. (A) Emergence of $-1$ defect over time at the 4-arm junction, visualized as POM image ($\mathrm{Er} = 8$). (B) Polarization micorgraphs of the nematic flow texture show the transition from a defect-free state to a $-1$ topological defect at the junction ($\mathrm{Er}\approx 2$). The defect can stretch at higher Er, shown here at $\mathrm{Er}\gtrsim5$. (C) Numerically simulated nematic director field with increasing Er within the plane indicated in the inset; director is shown in blue and defects as isosurfaces of nematic degree of order $S=0.4$.
%The inset shows position and orientation of the cross-section at which the director (blue rods) and the scalar order parameter (isosurface at $S=0.4$) are plotted. Upper panel shows marginal reorientation of the director along the channel length at the maximum speed in the channel, $v_\text{max}=0.11\,\xi_N/\tau_N$. At $v_\text{max}=0.29\,\xi_N/\tau_N$, a small defect loop forms at the junction center once the flow-aligned director field has been established (middle panel), and a vertically stretched defect loop is emerges at higher flow speeds (lower panel). 
(D, E) Continuous, defect-free director field observed in specific inflow-outflow combinations, in experiments (upper panel) and confirmed in simulations (middle and lower panels). (F, G) When symmetrical flow conditions are restored (each inflow arm is flanked by 2 outflow arms), a $-1$ defect emerges, and stabilizes at the geometric center of the $4-$arm junction when the flow speed in all the arms are equal (G).}
\end{figure*}

\section{\label{sec:defect_unbinding}Charge Fractionalization and Defects Unbinding}

Defects having large negative topological charge (i.e. $k<-1$), are prone to decay into multiples of $-1$-defects. We have experimentally resolved the dynamics of the collapse of the defect loop at the central junction. Figure \ref{fig:2}A (lower panel) shows POM micrographs of the defect loops immediately after their formation. At the center of the $4-$arm junction, we observe a defect loop of charge $-1$ (Fig. \ref{fig:2}A lower left panel), which within a short time stabilizes into a $-1$ monopole of the pseudo-planar texture \cite{Pieranski:2016}. The defects in the $6-$ and the $8-$arm junctions, emerge as loops of charge $-2$ and $-3$ respectively (Fig. \ref{fig:2}A lower middle and right panels), and gradually decay into multiple $-1$-charged defects (Fig. \ref{fig:2}B). As presented in Fig. \ref{fig:2}B (upper panel), the $-2$ loop fractionalizes into two smaller $-1$ loops, and within $10$ s, stabilized into a pair of $-1$ defects. The fractionalization of the $-3$ loop (Fig. \ref{fig:2}B, lower panel) proceeds in three steps: {\em 1)} First, a loop of charge $-3$ splits into a $-2$ loop and a $-1$ loop. {\em 2)} The $-1$ loop shrinks, while the $-2$ loop splits into two $-1$ loops. {\em 3)} Finally, all three $-1$ defect loops shrink down to the $-1$ structure, completing the fractionalization process. These emergent $-1$ monopoles are singularities of the pseudo-planar texture, whose positions are stable over time. However, their relative arrangement can be changed by tuning the flow within arms of the junction (Appendix \ref{sec:experimental_setup}, Appendix Fig. \ref{fig:multidyn}). 

The behavior described above results from two competing effects. On the one hand, the hydrodynamic forces tend to concentrate the negative topological charge at the center of the junction. On the other hand, the elastic forces drive the repulsion of like-sign defects. This favors the fractionalization of a central $k=-n$ topological charge, into $n$ defects of charge $-1$. Furthermore, hydrodynamic stagnation points of charge $-2$ and $-3$ (Fig. \ref{fig:1}C), are susceptible to decay, and can become unstable with respect to any perturbation of the pressure distribution across the channels. A slight asymmetry in the pressure distribution causes the central stagnation point to split into multiples of stagnation points of charge $-1$, thus further favoring the unbinding of defects.

\section{Dynamics of Defect Nucleation in a $4-$arm Junction}

Upon starting the flow synchronously in a 5CB-filled $4-$ arm microfluidic device, the director field aligns along the flow direction. The alignment initiates close the respective inlets of the opposite-facing arms, however, further downstream, the director field remains relatively undisturbed. Thus, each inlet arm devolps two director domains: upstream, a flow-aligned director domain; and downstream, an unperturbed homeotropic domain. These two director domains are separated by a disclination line with winding number $1/2$ \cite{Sengupta:2013b,Sengupta:2012b}. The disclination travels downstream in each of the facing inflow arms (Appendix Fig. \ref{fig:nucleation}), and meet head-on at the junction center (Fig. \ref{fig:3}A, middle panel). Upon meeting, the singular disclinations merge into a defect loop, enclosing a homeotropic domain (Fig. \ref{fig:3}A, fourth panel from the top), which gradually shrinks, and finally stabilizes into a $-1$ defect at the junction center (Appendix \ref{sec:new}). We would like to emphasize that homeotropic anchoring, in absence of flow, supports multiple director configurations. These energetically stable or metastable configurations emerge due to an interplay between the cross-section geometry (rectangular, square or circular), anchoring strength, and the curvature (or sharpness) of the channel corners \cite{Sengupta:2014,Sengupta:2013b}, and set the initial conditions for our flow experiments.

In a second approach, we have gradually increased the flow speed (in steps of Er = $0.5$) in each inflow arm, and allow the director field to equilibrate before increasing Er further. The exact structure of nematic field, and the emergence of the nematic topological defects are observed to be strongly dependent on the Ericksen number, which we vary by changing the magnitude of the flow field. Fig. \ref{fig:3}B presents sequence of polarized micrographs of the nematic texture at the junction center. The first appearance of the $-1$ defect loop was recorded at $\mathrm{Er}\approx 2$. At higher Er, $-1$ defect loop was located stably at the center, however, could extend along one or either sides of the outflow arms (Fig. \ref{fig:3}B, panels 4 and 5 from top). %We confirm the defect creation using numerical simulation. As shown in Fig. \ref{fig:3}C, at low Er values, the director orientation in the channel is determined by the surface anchoring. 
The profile of the director within the $4-$arm junction is obtained by using numerical modeling (see Fig.~\ref{fig:3}C). Increasing the flow speed (or Er) results in a further pronounced flow-alignment of the director, and at still larger Er values, the system attains a complete flow-alignment with the nematic director aligned roughly parallel to the channel direction. As the two flow-aligned domains meet at the junction center, the mismatch in the nematic director leads to the formation of a small defective loop of charge $-1$ (Fig.~\ref{fig:3}C, middle panel). At high Er values, the flow forces take over the elastic forces, and determine the director field in the proximity of the newly emerged nematic defect \cite{Sengupta:2014}. As a consequence of viscous forces, the defect loop can also flip and stretch out of the vertical plane (Fig.~\ref{fig:3}C, bottom panel). A stable $-1$ defect loop can also emerge by designing a specific modulation of the flow at the $4-$arm junction. As shown in Fig. \ref{fig:3}D, a combination of $3$ inflow arms (left, right and top), and $1$ outflow arm (bottom), results in a defect-free state at the junction center. By switching off the inflow in the top arm (Fig.~\ref{fig:3}E), the system gradually reorganizes and, as symmetric outflow conditions restore, a transition to the defective configuration (Fig.~\ref{fig:3}F-G) is observed. This result shows that by designing different microfluidic circuits -- and junction geometries -- could be used as an interesting route for creation of nematic defect structures of various complexity. 

\begin{figure*}[t]
\centering
\includegraphics[width=\textwidth]{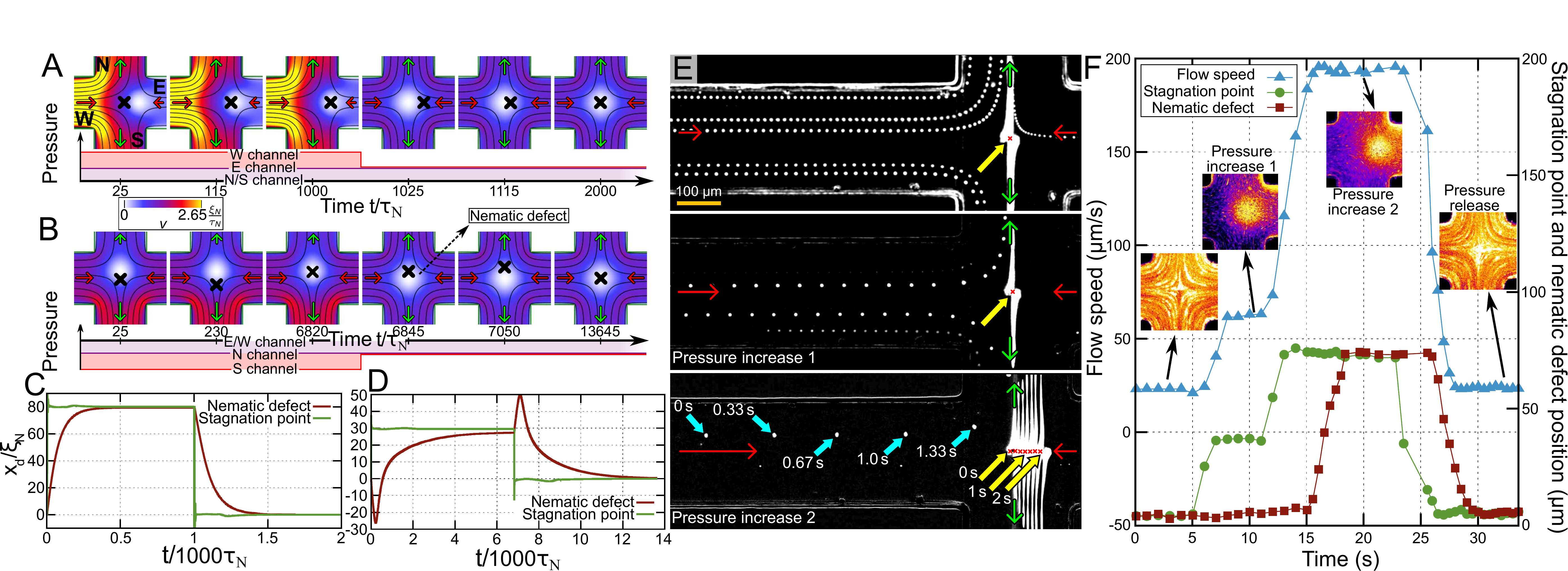}
\caption{\label{fig:4}Cross-talk between topological defects in different fields. (A) Simulations show displacement of the nematic defect and the hydrodynamic stagnation point when pressure in the left arm (W) of device was increased. The stagnation point shifts to a new position, followed by a much slower shift of the nematic defect. Before the pressure pulse is turned off (panel 3 from left), the stagnation point and the nematic defect are completely re-aligned. After the pressure is released, the stagnation point shifts back to the original position, slowly followed by the nematic defect. The exact position of the stagnation point and the nematic defect over time is shown in (C). (B) When pressure in the bottom arm is decreased, the nematic defect first drifts away from the shifted stagnation point (i.e., against the flow). At longer times the nematic defect approaches the stagnation point, and finally they overlay (D). (E) Positions of the hydrodynamic stagnation point and the nematic defect are measured experimentally as function of time. White dots show the transport of tracer particles over time, and the yellow arrow head indicates the position of the nematic defect center. Insets in (F) show the corresponding positions of the hydrodynamic stagnation point over time, obtained from the fluorescence measurements of the tracer particle flow at the junction center. (F) The increased pressure pulse in experiments confirm numerical results: the stagnation point first undergoes an instantaneous shift, and then the nematic defect drifts toward the stagnation point. Upon releasing the pressure, the defects return back to their initial position, starting with the stagnation point, and then followed by the nematic defect.}
\end{figure*}

\section{Discussions}

The coupling between the velocity and the orientational fields serves as a tunable mechanism for designing multi-field topology in nematic microfluidic systems. Our results reveal that this coupling underpins also the cross-interaction between the topological defects in the velocity and the nematic orientational fields. We quantify the strength of the interaction between the hydrodynamic and the nematic defects by perturbing the defects out of their equilibrium position in a $4-$arm junction, and analyzing the relative shift between the defects as a function of time. Altering the inlet pressure in one of the flow arms displaces the stagnation point off the center first, followed by gradual recovery of the nematic director. Fig. \ref{fig:4}A,C presents this dynamics using numerical calculations dynamics. Once the stagnation point and the nematic defect are separated (Fig. \ref{fig:4}A, left panel), the latter approaches the stagnation point and, within $1000\tau_N$,
%in a time $10^{3}$ times longer than the characteristic relaxation time scale of the nematic phase, 
the stagnation point and the nematic defect coincide again. Altering the pressure in an outflow arm also shifts the stagnation point first, followed by recovery of the $-1$ defect (Fig. \ref{fig:4}B,D). However, as the nematic defect now moves against the flow, the recovery is 10 times slower than in the previous case. Furthermore, the nematic defect initially moves backward before progressing toward the hydrodynamic stagnation point at the new location.

Experimentally, we perturb our system out of the equilibrium state, by marginally increasing the inlet pressure in the left arm (Fig. \ref{fig:4}E), and record the position of the defects over time. By overlaying consecutive frames of the recorded video data, we obtain a processed micrograph that captures the transport of tracer particles (bright dots along the flow direction), and the position of the defect over time (indicated by the yellow arrow head). The separation between bright dots is the distance traveled by a particle over the time interval between consecutive frames. This gives us a flow speed of $24$ $\mu$m/s under equilibrium conditions, as shown in Fig. \ref{fig:4}E (top panel). The topological defects remain co-localized at the center of the junction (no relative shift, 0-5 s in Fig. \ref{fig:4}F). When we increase the inlet pressure in the left arm (\ref{fig:4}E middle panel), the flow speed increases to $\approx 62$ $\mu$m/s, and shifts the stagnation point (Pressure increase 1 $\approx 15$ kPa, Fig. \ref{fig:4}F) by $\approx 40$ $\mu$m to the right. The $-1$ nematic defect however remains locked at the center of the junction. Only upon increasing the pressure further ($v = 180$ $\mu$m/s), the nematic defect shifts. As shown in Fig. \ref{fig:4}E (bottom panel), the defect shifted by $\approx90$ $\mu$m, before finally coinciding with the stagnation point at the new equilibrium position (Pressure increase 2 $\approx 50$ kPa, Fig. \ref{fig:4}F). When the perturbing pressure was released, the stagnation point rapidly returned to the junction center, followed slowly by the $-1$ defect (Pressure release in Fig. \ref{fig:4}F). The observed dynamics demonstrates a complex interaction between the hydrodynamic stagnation point and the nematic defect, which is clearly dependent on the direction of motion of the nematic defect relative to the local material flow. More generally, and in a mechanics-motivated view, the emergent dynamics of the two defect types in the vicinity of each other, could be viewed as induced by an inter-defect force (or potential) that stems from the coupling of the two material fields; and is inherently mediated by the topology (i.e., the topological charge) of the involved defects.

The cross-interaction between topological defects originating from different fields, though demonstrated in the context of nematic microfluidics, is a phenomenon, which, owing to its topological nature, is much more general in appeal. The demonstrated cross-talk relies on the existence of multiple spatially overlying material fields -- in our case vector-type, but could also be scalar or tensorial -- which are mutually coupled by some force-, stress- or energy-like cross-coupling mechanism. Therefore, the natural candidates for such phenomena will be systems with pronounced transport effects, or strongly interacting field. As possibly the most far reaching question of this type, such concepts of cross-field interacting defects could offer a physical framework for addressing phenomena in systems as diverse as cosmology, where objects like black holes are known singularities in the continuum of space and time.

In conclusion, the interplay between fluid flow and molecular orientation in nematic microfluidics has revealed that a hydrodynamic stagnation point can create a nematic defect of same topology, and that their strengths - both integer and semi-integer (Appendix \ref{sec:6}) can be tuned hierarchically, using microfluidic geometry of star-shaped junctions with various combinations of flow inlets and outlets. Importantly, our experiments, numerical modeling, and analytical calculations, demonstrate that topological defects in different material fields cross-talk, and their characterization reveals a unusual topologically-conditioned
% strength. To the best of our knowledge, this is the first report on  
interaction between these defects of hydrodynamic and nematic-ordering origin. Given that topological defects of disparate origin coexist in a range of physical and biological matter, this work could introduce a fresh perspective towards exploring and designing novel
%will serve as a first step toward elucidating the role of singularities of one type on the static and dynamical properties of the other type, 
%or if they can be harnessed to modulate attributes of the other type, and highlight explored 
material systems underpinned by multi-field topological defects. 

\section{Materials and Methods}

\subsection*{Experimental Setup}

We have used 4$^\prime$-pentyl-4-biphenylcarbo-nitrile (5CB), a single component nematic LC ($18^{\circ}\mathrm{C}<T<33^{\circ}\mathrm{C}$) for experiments. The microfluidic channels had rectangular cross-section, with depth $d~\approx 10$ $\mu$m, width $w = 100$ $\mu$m, and 15 mm length (unless otherwise specified). The channels were treated with $0.1\%$ w/w aqueous solution of silane DMOAP to create homeotropic surface anchoring \cite{Sengupta:2014}. Prior to flow experiments, microchannels were filled with 5CB in its isotropic phase. Once cooled down to the nematic phase, we have gradually increased the flow rate till topological defects emerged at the channel junction. The flow rate was varied between 0.01 - 2.0 $\mu\mathrm{l}/\mathrm{h}$ (corresponding flow speed, $v$, ranged between $2~\mu\mathrm{m}/\mathrm{s}$ and $0.40\,~\mathrm{mm}/\mathrm{s}$) in each arm. Thus, the characteristic Reynolds number $\mathrm{Re} = \rho vl/\eta$ ranged between $10^{-6}$ and $10^{-4}$. Here, $\eta = 50\,~\mathrm{mPa\,s}$ and $\rho = 1025\,~\mathrm{kg/m^{3}}$ are the dynamic viscosity and density of 5CB, and $l = 4wd/2(w+d) \approx 18\,~\mu\mathrm{m}$ is the hydraulic diameter of the rectangular microchannels. The corresponding Ericksen number, $\mathrm{Er} = \eta v l/K$, $K = 5.5$ pN being the 5CB elastic constant (one-constant approximation), varied between 0.3 and 65.

\subsection*{Numerical Simulations}

Our numerical simulations rely on Beris-Edwards formulation of nematodynamics~\cite{Beris:1994} describing the evolution of system density, velocity, and nematic tensor order parameter by the coupled continuity equation, Navier-Stokes equation, and Beris-Edwards equation. Coupling between flow and orientational order is included by the nematic stress tensor and by the flow-driven deformations of the nematic tensor order parameter profile that compete with the relaxation of nematic orientation towards the free energy minimum. The nematic free energy is constructed phenomenologically, including terms describing phase behavior, effective elasticity, and surface anchoring~\cite{DeGennes:1995}. Continuity and Navier-Stokes equations are solved numerically by a lattice Boltzmann algorithm~\cite{Denniston:2001b} with open boundaries and pressure driven flows through the channels. Simultaneously, evolution of nematic tensor order parameter -given by the Beris Edwards equation- is solved by a finite difference algorithm. Further details on the model, applied hybrid numerical scheme, and the numerical parameters used is given in Supplementary Information (Appendix \ref{sec:numerical_simulations}).

\acknowledgments

LG is supported by The Netherlands Organization for Scientific Research (NWO/OCW). MR and ZK by the Slovenian Research Agency ARRS, through Grants No. J1-7300, L1-8135 and P1-0099, and US AFOSR EOARD grant FA9550-15-1-0418 (contract no. 15IOE028). AS thanks Human Frontier Science Program Cross Disciplinary Fellowship (LT000993/2014-C) for support; Stephan Herminghaus and Christian Bahr for discussions at different stages of this work; and the Max Planck Society for funding the initial phase of this work at the Max Planck Institute for Dynamics and Self-Organization (MPIDS), G{\"o}ttingen, Germany. The authors thank Simon Čopar for insightful discussions on the dynamics of defect nucleation.

\appendix

\section{\label{sec:experimental_setup}Experimental Setup and Microfluidic Manipulations}

\begin{figure}[t]
\centering
\includegraphics[width= 1.0\columnwidth]{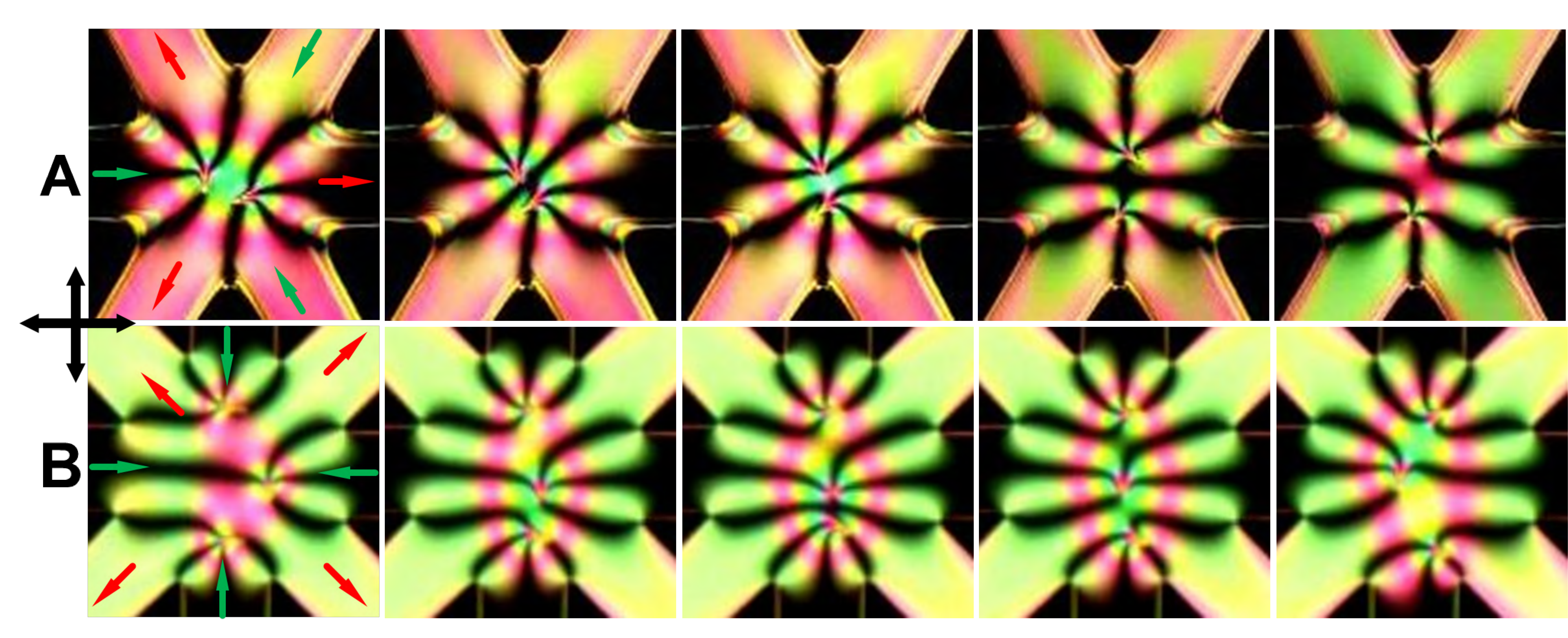}
\caption{\label{fig:multidyn} Hydrodynamic stagnation as a topological template for generation of defects with defined effective strengths. Polarized optical micrographs show arrangement of multiple $-1$ defects constituting an effective topological charge of –2 (A) and –3 (B) at the center of a $6-$ and $8-$arm channel respectively. The crossed-polarizers are indicated by the double-headed arrows (black). The relative position of the $-1$ defects could be rearranged by perturbing the flow field.}
\end{figure}

All our experiments have been performed with 4$^\prime$-pentyl-4-biphenylcarbo-nitrile, commonly known as 5CB (Synthon Chemicals). This is single component nematic liquid crystal for $18^{\circ}\mathrm{C}<T<33^{\circ}\mathrm{C}$ and was used without any additional purification. The arms of the microfluidic devices have rectangular cross-section, with depth $d~\approx 10$ $\mu$m and width $w = 100$ $\mu$m (unless otherwise specified). The length of each arm (15 mm), is much larger compared to the other two dimensions. The walls of the microfluidic arms were chemically treated with an aqueous solution of octadecyldimethyl(3-trimethoxysilylpropyl)ammonium chloride (DMOAP) to create a strong, homogeneous homeotropic surface anchoring. The channel was first filled with the DMOAP solution and then rinsed with deionized water (after $\approx 10$ min), after which the anchoring conditions within the channels were stabilized by thermal treatment at $80^{\circ}$C for 15 min and at $110^{\circ}$C for 1 h. This yields homogeneous homeotropic surface anchoring conditions on all the surfaces. Our microfluidic devices were first filled with 5CB in the isotropic phase, and allowed to cool down to nematic phase at room temperature. Thereafter, we progressively increased the volume flow rate to observe the first appearance of the topological defects at the junction center. We varied the flow rate in the range $[0.01,2.0]$ $\mu\mathrm{l}/\mathrm{h}$  corresponding to a flow speed $v$ ranging between $\approx 2\,~\mu\mathrm{m}/\mathrm{s}$ and $0.40\,~\mathrm{mm}/\mathrm{s}$ in each arm. Thus, for 5CB having bulk dynamic viscosity, $\eta\approx 50\,~\mathrm{mPa\,s}$ \cite{Sengupta:2011}, the characteristic Reynolds number $\mathrm{Re} = \rho vl/\eta$ ranged between $10^{-6}$ and $10^{-4}$. Here, $\rho\approx 1025\,~\mathrm{kg/m^{3}}$ is the material density, and $l = 4wd/2(w+d) \approx 18\,~\mu\mathrm{m}$ is the hydraulic diameter of the rectangular microchannels. The specific geometry of the higher strength defects (Fig. \ref{fig:multidyn}) was manipulated by adjusting the hydrodynamic flow using inbuilt flow profile routines of the microfluidic pumps used for this work (neMESYS, Cetoni GmbH, Germany). 

The hydrodynamic stagnation point in each experiment was detected by epi-fluorescent video imaging of fluorescent tracers (mean diameter 2.5 $\mu$m, $\lambda_{ex}$ = 506 nm, $\lambda_{em}$ = 529 nm) dispersed in the flowing NLC. As the particles approached the vicinity of the stagnation point, their speed diminished, and consequently, the residence time increased. Upon averaging the fluorescent intensity over multiple frames of the acquired video micrograph, the stagnation point appeared as a high intensity (bright) spot, relative to the surrounding region, due to the increased residence time of the particles in the stagnation point. For each experiment we toggled the microscopy modes (between epi-fluorescent and polarization optical microscopy) at quick successions ($\approx$ 0.5 s) to identify the corresponding position of the topological defects. Alongside, the fluorescent particles also served as tracers for the flow velocity measurements. The video micrographs were analyzed using a standard routine for tracking and trajectory analysis available through MATLAB. By keeping the tracer concentration very low, and by sonicating the dispersion freshly before each experiment, we ensured that the tracer particles did not self-assemble into bigger clusters in our experiments. 

\section{\label{sec:numerical_simulations}Numerical Simulations}

Our numerical simulations rely on Beris-Edwards formulation of nematodynamics \cite{Beris:1994} describing the evolution of the system by density $\rho$, velocity $\bm{v}$ and full order parameter nematic tensor $\bm{Q}$:
\begin{subequations}\label{eq:beris_edwards}
\begin{align}
%\partial_t\rho+\partial_k\left(\rho v_k\right)=0,\\
%\rho\left(\partial_t+v_k\partial_k\right)v_i=\partial_j\Pi_{ij},\\
%\left(\partial_t+v_k\partial_k\right)Q_{ij}-S_{ij}=\Gamma H_{ij}.
&\partial_{t}\rho+\nabla\cdot(\rho\bm{v}) = 0,\\[5pt]
&\rho(\partial_{t}+\bm{v}\cdot\nabla)\bm{v} = \nabla \cdot \bm{\sigma},\\[5pt]
&(\partial_{t}+\bm{v}\cdot\nabla)\bm{Q}-\bm{S}=\Gamma \bm{H}.
\end{align}
\end{subequations}
The dynamics of the nematic tensor is governed by the ``molecular field'' tensor $\bm{H}$:
\begin{equation}
%H_{ij}=-\frac{\delta \mathscr{F}_{\rm LdG}}{\delta Q_{ij}}+\frac{1}{3}\mathrm{Tr}\left(\frac{\delta \mathscr{F}_{\rm LdG}}{\delta Q_{ij}}\right)\delta_{ij}\;,
\bm{H}=-\frac{\delta \mathscr{F}_{\rm LdG}}{\delta \bm{Q}}+\frac{1}{3}\mathrm{Tr}\left(\frac{\delta \mathscr{F}_{\rm LdG}}{\delta\bm{Q}}\right)\bm{I}\;,
\end{equation}
where
\begin{multline}\label{eq:landau_degennes}
\mathscr{F}_{\rm LdG}
%= \frac{L}{2} \int \mathrm{d}V\,\partial_kQ_{ij}\partial_kQ_{ij}\\
%+ \int \mathrm{d}V\,\left[\frac{A}{2}Q_{ij}Q_{ji}+\frac{B}{3}Q_{ij}Q_{jk}Q_{ki}+\frac{C}{4}(Q_{ij}Q_{ji})^2\right] \\
%+ \frac{W}{2} \int \mathrm{d}A\,\mathrm{Tr}\left(Q_{ij}-Q_{ij}^0\right)^2
= \int \mathrm{d}V\,\left[\frac{A}{2}\mathrm{Tr}\,\bm{Q}^{2}+\frac{B}{3}\mathrm{Tr}\,\bm{Q}^{3}+\frac{C}{4}(\mathrm{Tr}\,\bm{Q}^{2})^2\right] \\[5pt]
+ \frac{L}{2} \int \mathrm{d}V\,|\nabla\bm{Q}|^{2} + \frac{W}{2} \int \mathrm{d}A\,\mathrm{Tr}\left(\bm{Q}-\bm{Q}_{0}\right)^2
\end{multline}
is the Landau-de Gennes free energy \cite{DeGennes:1995} augmented by the surface anchoring energy \cite{Nobili:1992} with preferred nematic tensor $\bm{Q}_{0}$. The tensor
\begin{align}
%\begin{multline}
%S_{ij}
%&=
%(\xi u_{ik}+\omega_{ik})\left(Q_{kj}+\frac{\delta_{kj}}{3}\right) \notag\\
%&+\left(Q_{ik}+\frac{\delta_{ik}}{2}\right)(\xi u_{kj}-\Omega_{kj}) \notag\\
%&-\frac{4}{3}\xi\left(Q_{ij}+\frac{\delta_{ij}}{2}\right)Q_{kl}partial_{l}v_{k},
\bm{S}
&= (\xi\bm{u}-\bm{\omega})\cdot\left(\bm{Q}+\frac{1}{3}\bm{I}\right)\notag\\[5pt]
&+ \left(\bm{Q}+\frac{1}{3}\bm{I}\right)\cdot(\xi\bm{u}+\bm{\omega})\notag\\[5pt]
&-2 \xi \left(\bm{Q}+\frac{1}{3}\bm{I}\right)\mathrm{Tr}(\bm{Q}\cdot\nabla\bm{v}),
\end{align}
%\end{multline}
with $u_{ij}=(\partial_{i}v_{j}+\partial_{j}v_{i})/2$ and $\omega_{ij}=(\partial_{i}v_{j}-\partial_{j}v_{i})/2$ the strain-rate and vorticity tensor, embodies the interaction between local orientation and flow. Finally, the stress tensor $\bm{\sigma}$ is given by:
%$\Pi_{ij}=-P\delta_{ij}+\eta(\partial_iv_j+\partial_jv_i)+\Sigma_{ij}$, with $P=P_{0}-(L/2)\partial_kQ_{ij}\partial_{k}Q_{ij}$ the total pressure and
%\begin{multline}
%\Sigma_{ij}=
%Q_{ik}H_{kj}-H_{ik}Q_{kj}
%-\partial_iQ_{kl}\frac{\delta F}{\delta\partial_jQ_{kl}}
%-\lambda H_{ik}\left(Q_{kj}+\frac{\delta_{kj}}{2}\right)\\
%-\lambda\left(Q_{ik}+\frac{\delta_{ik}}{2}\right)H_{kj}
%+\frac{4}{3}\lambda\left(Q_{ij}+\frac{\delta_{ij}}{2}\right)Q_{kl}H_{lk},
%\end{multline}
\begin{align}
\label{eq:BE_sigma}
\bm{\sigma}
&= - P\bm{I}+2\eta\bm{u}\notag\\[5pt]
&- \xi \bm{H}\cdot\left(\bm{Q}+\frac{1}{3}\bm{I}\right)    
 - \xi \left(\bm{Q}+\frac{1}{3}\bm{I}\right)\cdot\bm{H}\notag\\[5pt]
&+2\xi \left(\bm{Q}+\frac{1}{3}\bm{I}\right) \mathrm{Tr}(\bm{Q}\cdot\bm{H}) \notag\\[5pt]
&+ \bm{Q}\cdot\bm{H}-\bm{H}\cdot\bm{Q}+\bm{\sigma}^{\rm E},
\end{align}
where:
\begin{equation}
P = P_{0}-\frac{L}{2}|\nabla\bm{Q}|^{2}   
\end{equation}
is the pressure and:
\begin{equation}
\sigma^{\rm E}_{ij} = -\partial_{j}Q_{kl}\frac{\delta\mathscr{F}_{\rm LdG}}{\delta\partial_{i}Q_{kl}}   
\end{equation}
is the Ericksen stress.  The following parameter values are used in our simulations: $A=-0.172\times10^6\,\text{Jm}^{-3}$, $B=-2.12\times10^6\,\text{Jm}^{-3}$, $C=1.73\times10^6\,\text{Jm}^{-3}$, $L=4\times10^{-11}\,\text{N}$, $W=10^{-2}\,\text{J/m}^2$, $\Gamma=16\,\text{Pa}^{-1}\,\text{s}^{-1}$, $\xi=0.94$. This yields the following values of the nematic order parameter $S_{0}=0.533$, nematic correlation length $\xi_N=[L/\left(A+BS_0+9CS_0^2/2\right)]^{1/2}=6.63\,\text{nm}$ and $\tau_N=\xi_N^2/(\Gamma L)=6.8\times10^{-8}\,\text{s}$. \eqref{eq:beris_edwards} were numerically integrated using the hybrid lattice Boltzmann method \cite{Denniston:2001,Sengupta:2013} with a 19 velocity lattice model and Bhatnagar-Gross-Krook collision operator \cite{Succi:2001}. The flow was driven by a pressure difference with an open boundary at the end of the channels. In the studied regime, the fluid is nearly incompressible and small density gradients (the deviations of density in a junction are always kept below $\approx 2\,\%$) are only used to induce pressure difference in microchannels with $P_0$ taken to be proportional to the density. At average density, $\eta$ in \eqref{eq:BE_sigma} evaluates to $78\,\text{mPas}$. Resolution of the numerical mesh is set to $1.5\xi_N=10\,\text{nm}$, which still ensures that there is no pinning of the nematic defects to the mesh points~\cite{Ravnik:2009}. In case of a two-dimensional system of uniform density and nematic order parameter, as used in our theoretical analysis, the full equations in \eqref{eq:beris_edwards} reduce to \eqref{eq:nematodynamics} with $\Gamma=9S^2/(2\gamma)$, $\xi=9 \lambda S/(3S+4)$, and $K=9LS^2/2$.

\section{\label{sec:nematodynamics}Flow-alignment in Nematics}

In this section we provide additional information on the hydrodynamics of nematic liquid crystals as well as a derivation of Eq. (1) in the main text. In the presence of uniform density and nematic order (i.e. $\rho,\,S=\mathrm{constant}$), the Beris-Edwards equations (\ref{eq:beris_edwards}), simplify to the following set of partial differential equations for the velocity field $\bm{v}$ and the nematic director $\bm{n}$ \cite{Landau:1986,DeGennes:1995,Chaikin:2000,Oswald:2005,Kleman:2007}:
\begin{subequations}\label{eq:nematodynamics}
\begin{gather}
\rho(\partial_{t}+\bm{v}\cdot\nabla)\bm{v} = \eta \nabla^{2}\bm{v}-\nabla p + \nabla\cdot\bm{\sigma}^{\rm el},\\[5pt]
(\partial_{t}+\bm{v}\cdot\nabla)\bm{n} + \bm{\omega}\cdot\bm{n}=\bm{\Pi}\cdot(\lambda\bm{u}\cdot\bm{n}+\gamma^{-1}\bm{h}), 
%(\partial_{t}+\bm{v}\cdot\nabla)n_{i} = -\omega_{ij}n_{j}+(\delta_{ij}-n_{i}n_{j})(\lambda u_{jk}n_{k}+\gamma^{-1}h_{j})\;,
\end{gather}	
\end{subequations}
where $u_{ij}=(\partial_{i}v_{j}+\partial_{j}v_{i})/2$ and $\omega_{ij}=(\partial_{i}v_{j}-\partial_{j}v_{i})/2$ are the strain-rate and vorticity tensor and $\Pi_{ij}=\delta_{ij}-n_{i}n_{j}$ is the transverse projection operator. The constants $\eta$ and $\gamma$ are the shear and rotational viscosity, while $\lambda$ is the flow-aligning parameter discussed in the main text. The relaxational dynamics of the nematic director is dictated by the molecular field $\bm{h}=-\delta\mathscr{F}_{\rm F}/\delta\bm{n}$ associated with the Frank free energy:
\begin{equation}\label{eq:frank}
\mathscr{F}_{\rm F} = \frac{1}{2}\int {\rm d}V\,\Big[K_{1}(\nabla\cdot\bm{n})^{2}+K_{2}(\bm{n}\cdot\nabla\times\bm{n})^{2}+K_{3}|\bm{n}\times\nabla\times\bm{n}|^{2}\Big],	
\end{equation}
where $K_{1}$, $K_{2}$ and $K_{3}$ are respectively the splay, twist and bending elastic constants. In one elastic constant approximation $K_{1}=K_{2}=K_{3}=K$ and the molecular tensor is $\bm{h}=K\nabla^{2}\bm{n}$. Finally, the elastic stress tensor is given by \cite{Landau:1986,Chaikin:2000}:
\begin{equation}\label{eq:elasticstress}
\sigma_{ij}^{\rm el} = -\frac{\lambda}{2}(n_{i}h_{j}^{\perp}+n_{j}h_{i}^{\perp})+\frac{1}{2}(n_{i}h_{j}^{\perp}-n_{j}h_{i}^{\perp}),
\end{equation}
where $\bm{h}^{\perp}=\bm{\Pi}\cdot\bm{h}$. In all our analytical calculations we have neglected variations in the direction perpendicular to the plane of the junction, thus rendering the problem effectively two-dimensional. As we anticipated in the main text, this assumption does not allow to capture the escaped structures observed in the numerical simulations,
however, does provide crucial insight to our understanding of the mechanisms governing the interaction between dynamical and topological defects.
%As we anticipated in the main text and we will see in more detail in Sec. \ref{sec:loops}, assumption does not allow to capture the escaped structure of the defect loops with topological charge $k \le -2$, but does provide crucial insight in our understanding the mechanisms governing the interaction between dynamical and topological defects. 
Taking $\bm{n}=(\cos\theta,\sin\theta,0)$, Eq. (\ref{eq:nematodynamics}b) can be cast in the form of a single partial differential equation for the local orientation $\theta$:
\begin{equation}\label{eq:theta}
(\partial_{t}+\bm{v}\cdot\nabla)\theta = \frac{K}{\gamma} \nabla^{2}\theta+\omega_{xy}-\lambda(u_{xx}\sin 2\theta-u_{xy}\cos 2\theta).	
\end{equation}
The flow aligning behavior of nematics becomes especially evident in the presence of a simple shear flow of the form $\bm{v}=(0,\dot{\epsilon}y,0)$, with $\dot{\epsilon}$ a constant shear-rate. \eqref{eq:theta} then reduces to:
\begin{equation}
(\partial_{t}+\bm{v}\cdot\nabla)\theta = \frac{K}{\gamma} \nabla^{2}\theta-\frac{\dot{\epsilon}}{2}(1-\lambda\cos 2\theta).
\end{equation}
Thus, sufficiently far from the boundary and for $\lambda \ge 1$, the director aligns at the equilibrium Leslie angle, $\theta=\arccos(1/\lambda)/2$ with respect to the flow direction \cite{DeGennes:1995,Oswald:2005}. Following up on the brief description after Eq. (\ref{eq:elasticstress}), we would like to note again that the analytical treatment presented here is strictly two-dimensional (variations in the direction perpendicular to the plane of the junction is neglected), and thus, cannot account for the escaped structures observed in our experiments and numerical simulations. Nevertheless, the two-dimensional analytical description serves as a relevant tool to gain insight to the mechanisms that underpin the interaction between the topological defects in different fields.

\section{\label{sec:flow}Stagnation Flows}

\begin{figure*}[t]
\centering
\includegraphics[width=0.9\textwidth]{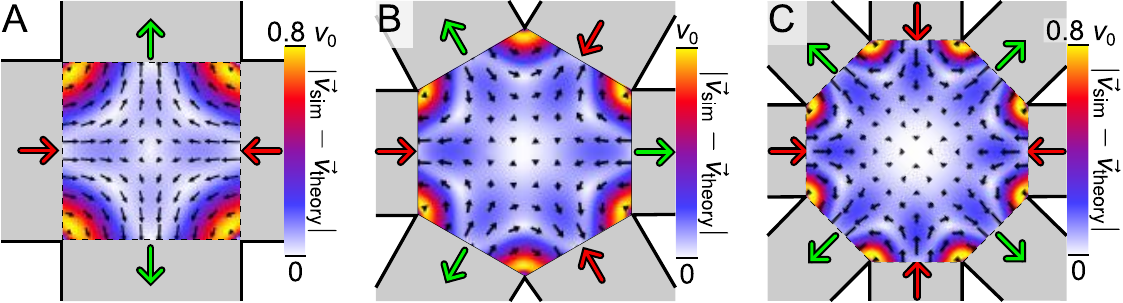}
\caption{\label{fig:comparison} Comparison between the approximated velocity field $\bm{v}_\text{theory}$, as given in Eqs. (\ref{eq:cross_flow_v}) and (\ref{eq:polygon_velocity}), and the simulated flow field $\bm{v}_\text{sim}$ at the horizontal cross section of the (A) 4-junction, (B) 6-junction, and (C) 8-junction. Results are rescaled to give the same flow rate through the channels on the cross-section. The approximated flow is in good agreement with the simulated one, except at the corner of the junction where the effects of viscous dissipations are dominant.}
\end{figure*}

The defective solutions and the force field reported in the main text have been constructed from analytic approximations of the stagnation flow in a $4-$, $6-$ and $8-$arm junction. Here we report an explicit construction of the corresponding velocity fields. Let $\bm{v}=(\partial_{y}\psi,-\partial_{x}\psi)$ be the two-dimensional velocity field in the mid-plane of the junction, with $\psi$ the associated stream function. In Eq. (\ref{eq:nematodynamics}a), the ratio between the magnitude $\eta v/l$ of viscous stresses, with $v$ the typical flow velocity and $l$ the system size, and the magnitude $K/l^{2}$ of elastic stresses, yields the Ericksen number $\mathrm{Er}=\eta v l/K$, whereas $\mathrm{Re}=\rho v l/\eta$ is the usual Reynolds number expressing the ratio between inertial and viscous force. For $\mathrm{Re}\approx 0$ and $\mathrm{Er} \gg 1$, both inertial and elastic effects can be neglected in Eq. (\ref{eq:nematodynamics}a) and the flow is governed by the incompressible Stokes equations:
\begin{equation}
\eta \nabla^{2}\bm{v}-\nabla p = \bm{0},\qquad
\nabla\cdot\bm{v} = 0.
\end{equation}
Consistently, the streamfunction $\psi$ obeys the biharmonic equation \cite{Batchelor:1967}:
\begin{equation}\label{eq:biharmonic}
\nabla^{4}\psi = 0.	
\end{equation}
Now, in order to construct an analytical approximation of the flow at the center of the junction, we look for the lowest order biharmonic stream function with the rotational symmetry of the junction and whose stagnation points are suitably located. As a starting point, let us consider a square domain of size $w$ representing the central region of a $4-$arm junction. Due to the symmetry of the inlets and outlets, the velocity field in the junction is characterized by the following symmetries:
\begin{subequations}\label{eq:symmetries}
\begin{align}
v_{x}(x,y) &=-v_{y}(y,x),\\[5pt]
v_{x}(x,y) &=-v_{x}(-x,y),
\end{align}
\end{subequations}
as well as those derived from these equations. Eq. (\ref{eq:symmetries}a), in particular, implies:
\begin{equation}
\psi(x,y) = \psi(y,x).	
\end{equation}
Consistently with this property, we can parametrize $\psi$ with two functions $f$ and $g$, such that:
\begin{equation}
\psi(x,y) = f(x)g(y)+f(y)g(x).
\end{equation}
By virtue of Eq. (\ref{eq:symmetries}b), both $f$ and $g$ must be odd functions, i.e. $f(x)=-f(-x)$, which in turn implies $f'$ to be an even function, i.e. $f'(x)=f'(-x)$. Following our calculation scheme, we choose $g(x)=x$ to obtain the lowest order stream function. The velocity field is then given by:
\begin{subequations}\label{eq:v_of_f}
\begin{align}
v_{x}(x,y) &= f(x)+xf'(y),\\[5pt]
v_{y}(x,y) &=-f(y)-yf'(x),
\end{align}
\end{subequations}
while the biharmonic equation can be cast into the form:
\begin{equation}\label{eq:f}
x\,\partial_{y}^{4}f(y)+y\,\partial_{x}^{4}f(x) = 0.
\end{equation}
Due to the separation of variables in \eqref{eq:f}, the function $f$ must be a third-order polynomial of the form:
\begin{equation}\label{eq:polynomial}
f(x) = a_{1}x+a_{3}x^{3}.
\end{equation}
The coefficients $a_{1}$ and $a_{3}$ can be determined by fixing the position of the stagnation points. The flow in a symmetric cross-junction with alternating inlets and outlets has, in fact, five stagnation points. One at the center of the junction and four at the corners, i.e. $(\pm w/2,\pm w/2)$,  due to the no-slip walls (Fig. \ref{fig:comparison}A). The flow described by Eqs. (\ref{eq:polynomial}) and (\ref{eq:v_of_f}) has at most 9 stagnation points, whose coordinates are given by:
\begin{subequations}
\begin{align}
\bm{r}_{0}^{\phantom{\pm}} &= (0,0),\\[10pt]
\bm{r}_{1}^{\pm} &= \left(0,\pm \sqrt{-\frac{2a_{1}}{a_{3}}}\;\right),\\[5pt]
\bm{r}_{2}^{\pm} &= \left(\pm \sqrt{-\frac{2a_{1}}{a_{3}}},0\;\right),\\[5pt]
\bm{r}_{3}^{\pm} &= \left(\pm \sqrt{-\frac{a_{1}}{2a_{3}}},\pm \sqrt{-\frac{a_{1}}{2a_{3}}}\;\right).
\end{align}
\end{subequations}
The four symmetric stagnation points at $\bm{r}_{3}^{\pm}$ are suitable to be mapped into the points at the corners of the junction. Thus, taking:
\begin{equation}
-\frac{a_{1}}{2a_{3}} = \left(\frac{w}{2}\right)^{2},
\end{equation}
we finally obtain the function $f$ in the form:
\begin{equation}\label{eq:f_final}
f(x) = \frac{1}{2} a_{3} x (w^{2}-2x^{2}).
\end{equation}
The last constant $a_{3}$, can be finally related with the maximal absolute velocity $v_{0}$ in the junction. This is attained by the flow at the center of the inlets/outlets, i.e. $(0,\pm w/2)$ and $(\pm w/2,0)$. Eqs. (\ref{eq:v_of_f}) and (\ref{eq:f_final}) yields $v_{0}=(3/8)w^{3}a_{3}$, from which we obtain:
\begin{equation}
\psi(x,y) = \frac{8v_{0}}{3w^{3}}\,xy(w^{2}-x^{2}-y^{2})\;.
\end{equation}
The corresponding velocity field is given by:
\begin{subequations}\label{eq:cross_flow_v}
\begin{align}	
v_{x}(x,y) &= \frac{8 v_{0}}{3w^{3}}\,x\,(w^{2}-x^{2}-3y^{2}),\\[5pt]
v_{y}(x,y) &=-\frac{8 v_{0}}{3w^{3}}\,y\,(w^{2}-y^{2}-3x^{2}).	
\end{align}
\end{subequations}
Fig. \ref{fig:comparison}A shows a comparison between the approximated velocity field and a numerical solution of the Stokes equation in a cross junction. The two agree closely, and fall always within a 10\% range, with an exception for the corners where viscous dissipation plays the dominating role.

Now, consistently with \eqref{eq:cross_flow_v}, the vorticity field obtained from the approximation describe here is the lowest order harmonic function with two-fold rotational symmetry:
\begin{equation}
\omega = \partial_{x}v_{y}-\partial_{y}v_{x} = \frac{32v_{0}}{w^{3}}\,xy \propto r^{2} \sin 2\phi. 
\end{equation}
This result can be generalized to construct a $n-$fold symmetric stagnation flows approximating the flows in symmetric $2n-$arm microfluidic junctions. Let us consider the following stream-function:
\begin{equation}
\psi =  Ar^{n}(1+Br^{2})\sin n\phi,
\end{equation}
where $A$ and $B$ are constants and $n\ge 2$ an integer. The corresponding vorticity is given by:
\begin{equation}
\omega = -\nabla^{2}\psi \propto r^{n} \sin n\phi.
\end{equation}
This is the lowest order harmonic function with $n-$fold rotational symmetry [i.e. $\omega(\phi)=\omega(\phi+2\pi/n)$]. As for the case of a cross-junction, the constants $A$ and $B$ can be adjusted in order to obtain the right positioning of the stagnation points and the maximal absolute velocity. Proceeding as in the previous case, one can obtain, after some algebra:
\begin{equation}\label{eq:polygon_stream_function}
\psi = \frac{v_{0}}{n}\,\frac{r^{n}}{\rho^{n-1}}\,\frac{R^{2}(n+2)-n r^{2}}{R^{2}(n+2)-n\rho^{2}}\,\sin n\phi,
\end{equation}
where $\rho=(w/2)\cot\pi/(2n)$ and $R=(w/2)\csc\pi/(2n)$ are respectively the inradius and the circumradius of the regular $2n-$sided polygon of edge length $w$ representing the center of the junction. The velocity field constructed from \eqref{eq:polygon_stream_function} vanishes at the corners of the polygon and is maximal in magnitude at the center of the edges. This can be conveniently verified using polar coordinates:
\begin{subequations}\label{eq:polygon_velocity}
\begin{align}
v_{r} &= v_{0}\,\left(\frac{r}{\rho}\right)^{n-1}\frac{R^{2}(n+2)-nr^{2}}{R^{2}(n+2)-n\rho^{2}}\,\cos n\phi, \\[5pt]
v_{\phi} &=-v_{0}\,\left(\frac{r}{\rho}\right)^{n-1}\frac{(n+2)(r^{2}-R^{2})}{R^{2}(n+2)-n\rho^{2}}\,\sin n\phi.
\end{align}
\end{subequations}
Now, at the corners of the junction:
\[
r=R,\qquad \phi = \frac{\pi}{n}\left(i+\frac{1}{2}\right),\qquad v_{r} = v_{\phi} = 0,
\]
with $i=1,\,2\ldots n$. On the other hand, at the center an inlet/outlet:
\[
r = \rho,\qquad \phi = \frac{\pi}{n}(i+1),\qquad v_{r} = v_{0}\cos \pi(i+1),\qquad v_{\phi} = 0\;. 
\]
A comparison between the velocity fields described by \eqref{eq:polygon_velocity} and those obtained from a numerical integration of the Naiver-Stokes equation are show in Fig. \ref{fig:comparison}C,D for the cases $n=3,\,4$.

Finally, sufficiently close to the central stagnation point $r\ll R$ and the stream function in \eqref{eq:polygon_stream_function} is approximated by the harmonic function:
\begin{equation}\label{eq:harmonic_psi}
\psi \approx \frac{v_{0}}{n}\,\frac{r^{n}}{\mathcal{R}^{n-1}}\sin n\phi,
\end{equation}
where the length scale $\mathcal{R}$ is given by:
\begin{equation}
\mathcal{R} = \rho \left[1-\frac{n}{n+2}\,\left(\frac{\rho}{R}\right)^{2}\right]^{\frac{1}{n-1}}.
\end{equation}
\eqref{eq:harmonic_psi} describes an irrotational flow (i.e. $\omega=0$), whose velocity field is given by:
\begin{subequations}\label{eq:velocity}
\begin{align}
v_{r} &= v_{0} \left(\frac{r}{\mathcal{R}}\right)^{n-1} \cos n\phi,\\[5pt]
v_{\phi} &= -v_{0} \left(\frac{r}{\mathcal{R}}\right)^{n-1}\sin n\phi,
\end{align}
\end{subequations}

\section{\label{sec:irrotational}Defect Configurations in Irrotational Flows}

We use nematic hydrodynamics to calculate the configuration of the nematic director in close proximity of the central stagnation point of a generic $2n-$arm junction. As we observed in Sec. \ref{sec:flow} and in the main text, here we have considered the limit of large Ericksen number so that the backflow effects are negligible (viscous interactions dominate over elastic ones). In close proximity of the center of the junction, the flow is approximatively irrotational with a velocity field given by Eqs. (\ref{eq:velocity}). The dynamics of a two-dimensional nematic director is governed by \eqref{eq:theta}. Because of the rotational symmetry of the problem it is convenient to work in polar coordinates. Then, expressing $\bm{n}=\cos\alpha\,\bm{\hat{r}}+\sin\alpha\,\bm{\hat{\phi}}$, with $\alpha=\theta-\phi$, allows to rewrite \eqref{eq:theta} as:
\begin{multline}\label{eq:alpha}
\partial_{t}\alpha+v_{r}\,\partial_{r}\alpha+\frac{v_{\phi}}{r}\,\partial_{\phi}\alpha+\frac{v_{\phi}}{r} \\[5pt]
= \frac{K}{\gamma}\,\nabla^{2} \alpha - \lambda ( u_{rr}\sin 2\alpha - u_{r\phi} \cos 2\alpha )\;.
\end{multline}
The strain-rates $u_{rr}$ and $u_{r\phi}$ can be straighforwardly calculated from \eqref{eq:velocity}:
\begin{subequations}\label{eq:strain_rate}
\begin{align}
u_{rr} &= -u_{\phi\phi} = (n-1)\,\frac{v_{0}}{\mathcal{R}}\,\left(\frac{r}{\mathcal{R}}\right)^{n-2}\cos n\phi,\\[5pt]
u_{r\phi} &= u_{\phi r} = -(n-1)\,\frac{v_{0}}{\mathcal{R}}\,\left(\frac{r}{\mathcal{R}}\right)^{n-2}\sin n\phi.
\end{align}	
\end{subequations}
In order to render \eqref{eq:alpha} dimensionless, we can rescale $r\rightarrow r/\mathcal{R}$, $t\rightarrow t/\tau_{\mathcal{R}}$, with $\tau_{\mathcal{R}}=\gamma \mathcal{R}^{2}/K$ the typical relaxational time the nematic director over the length scale $\mathcal{R}$. Thus, using \eqref{eq:strain_rate}, after some manipulation \eqref{eq:alpha} can be expressed in the dimensionless form:
\begin{multline}\label{eq:dimensionless}
\partial_{t}\alpha+r^{n-1}\cos n\phi\,\partial_{r}\alpha-r^{n-2}\sin n\phi\,(\partial_{\phi}\alpha+1)\\[5pt]
= \mathrm{De}^{-1}\nabla^{2}\alpha-\lambda(n-1)r^{n-2}\sin(2\alpha+n\phi)\;,	
\end{multline}
where $\mathrm{De}=\gamma v_{0} \mathcal{R}/K$ is the Deborah number expressing the product between the shear rate $v/\mathcal{R}$ and the relaxation time $\tau_{\mathcal{R}}$.

In spite of its strong nonlinearity, it is possible to find a family of stationary defective solutions of \eqref{eq:dimensionless} for specific values of the flow-alignment parameter $\lambda$, by constructing the minimizers of the Frank free energy having the same rotational symmetry of the imposed flow. To see this, let us consider an ideal defective configuration of strength $k$. In polar coordinates this is described by:
\begin{equation}\label{eq:ideal_solution}
\alpha = (k-1)\phi+\alpha_{0}\;.	
\end{equation}
Replacing this into \eqref{eq:dimensionless} we obtain:
\begin{equation}
k \sin n\phi = \lambda (n-1) \sin[(2k-2+n)\phi+\alpha_{0}]\;.
\end{equation}
This equation must hold for any $\phi$ value. Thus, setting without loss of generality $\alpha_{0}=0$, we obtain the following conditions for $k$ and $\lambda$:
\begin{equation}\label{eq:nk}
\left\{
\begin{array}{l}
n = \pm (2k-2+n)\;,\\[5pt]
k = \pm \lambda (n-1)\;.
\end{array}
\right.
\end{equation}
Choosing the positive sign, results into a single physical solution: 
\begin{equation}\label{eq:flow_tumbling}
k = 1\;,\qquad
\lambda = \frac{1}{n-1}\;.
\end{equation}
As $n \le 2$, $\lambda \le 1$, thus \eqref{eq:flow_tumbling} describes a special bulk configuration of the director in flow-tumbling nematics. Choosing the negative sign in \eqref{eq:nk}, on the other hand, yields a family of solutions with:
\begin{equation}\label{eq:flow_aligning}
k=1-n\;,\qquad
\lambda = 1\;.
\end{equation} 
\eqref{eq:flow_aligning} defines a set of defective configurations having $k<0$ and whose rotational symmetry is related to that of the flow field, namely:
\begin{equation}\label{eq:ideal_flow_aligning}
\alpha = -n\phi\;.
\end{equation}
Thus, in the presence of a cross-flow ($n=2$), a possible configuration consists of an isolated disclination of turning number $k=-1$ trapped by the flow at the center of the junction. For a hexagonal flow $(n=3)$, the central defects has turning number $k=-2$ and so on. These ideal defective configurations, however, only exist for perfectly flowing aligning nematics, for which $\lambda=1$. Although mathematically very special, this solution describes, at least approximatively, the majority of thermotropic nematic liquid crystals for which $\lambda \gtrsim 1$. In the case of the 5CB used in our experiment, $\lambda\approx 1.1$ \cite{Sengupta:2012b}. 

\section{\label{sec:defects}Defect Dynamics in a Flow}

In this section we provide a derivation of Eq. (2) in the main text. In the absence of backflow, the dynamics of the local orientation $\theta$, governed by \eqref{eq:theta}, can be thought as resulting solely form energy relaxation:
\begin{equation}\label{eq:relaxation}
\partial_{t}\theta = -\frac{1}{\gamma}\frac{\delta\mathscr{F}}{\delta \theta},
\end{equation}
where:
\begin{equation}\label{eq:potential_energy}
\mathscr{F} = \int {\rm d}A\, \left[\frac{1}{2}K|\nabla\theta|^{2}+U(\theta)\right],
\end{equation}
and $U(\theta)$ is a potential energy density, such that: 
\begin{equation}\label{eq:partial_U}
U'(\theta) = - \omega_{xy} + \lambda (u_{xx}\sin 2\theta - u_{xy} \cos 2\theta),	
\end{equation}
where the prime denotes partial differentiation with respect to $\theta$. We stress that such a description is possible here exclusively for $\mathrm{Er} \gg 1$. In this regime, the director is reoriented by the flow, while the latter is insensitive to the conformation of the director. The effect of the flow on the dynamics of the nematic director is then equivalent to that of a static external field. More generally, Eqs. (\ref{eq:relaxation}) and (\ref{eq:potential_energy}) can be used to describe the dynamics of the local orientation $\theta$ in the presence of any potential energy field, as that associated with an external magnetic or electric field.

\begin{figure}[t]
\centering
\includegraphics[width=\columnwidth]{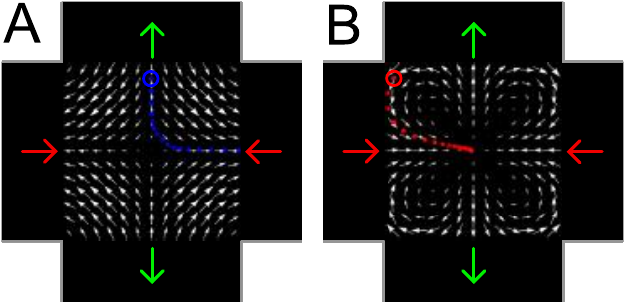}	
\caption{\label{fig:defect_force_field} The force field $\bm{F}$ experienced by a negative (left) and positive (right) disclination as a consequence of the flow, calculated using Eqs. (\ref{eq:partial_U}), (\ref{eq:force_flow}) and (\ref{eq:cross_flow_strain_rate}). The blue (red) dots represents the trajectory of a negative (positive) disclination starting from a generic circled point in the square region at the center of the junction. Negative defects are attracted by the central stagnation point, while positive disclinations are repelled toward the channels.}
\end{figure}

Let $\bm{R}=(X,Y)$ be the position of an isolated defect of topological charge $k$, traveling across the system as dictated by \eqref{eq:relaxation}. Following Kawasaki \cite{Kawasaki:1984} and Denniston \cite{Denniston:1996}, one can construct an equation of motion for the moving defect by decomposing the local orientation $\theta$ as:
\begin{equation}\label{eq:theta_decomposition}
\theta(\bm{r},\bm{R}) = \theta_{\rm d}(\bm{r},\bm{R})+\theta_{\rm ext}(\bm{r}).
\end{equation}
The field $\theta_{\rm d}$ describes the orientation of the director in the neighborhood of the defect core and is such that:
\begin{equation}\label{eq:theta_defect}
\theta_{\rm d}(\bm{r},\bm{R}) \xrightarrow{\bm{r}\rightarrow\bm{R}} k\arctan\left(\frac{y-Y}{x-X}\right),
\end{equation}
whereas $\theta_{\rm ext}$ represents the departure from this configuration away from the core. We would like to note here that the decomposition given in Eq. (\ref{eq:theta_decomposition}), results from the special structure of the director field near the core region, and does not require the linearity of the associated field equation \cite{Kawasaki:1984}. In order to find an equation of motion relating $\bm{R}$, with $\theta_{\rm d}$ and $\theta_{\rm ext}$, we calculate the energy variation due to a small virtual displacement $\delta\bm{R}$ of the defect:
\begin{equation}\label{eq:integral_eom}
\int {\rm d}A\,\delta\theta\,\partial_{t}\theta = -\frac{1}{\gamma}\,\delta \mathscr{F},	
\end{equation}
where $\delta\theta$ represent the variation in the director orientation caused by the defect displacement and the integral is performed over a punctured domain which excludes the defect core. Now, the energy variation $\delta \mathscr{F}$ in \eqref{eq:integral_eom} consists of a combination of a bulk term and a boundary term due to the shift in the position of the finite size core region. Namely:
\begin{multline}\label{eq:delta_F1}
\delta \mathscr{F} 
= \int {\rm d}A\, \left[ K \nabla \theta \cdot \nabla\delta\theta + U'(\theta)\delta\theta\right]\\
+ \oint {\rm d}s\,\delta\bm{R} \cdot \bm{N} \left[\frac{1}{2}K\,|\nabla\theta|^{2}+U(\theta)\right],
\end{multline}
where $\bm{N}$ is the boundary normal pointing toward the interior of the defect core. The variation $\delta\theta$ due to the defect displacement can by straightforwardly calculated from Eqs. (\ref{eq:theta_decomposition}) and (\ref{eq:theta_defect}) in the form:
\begin{equation}\label{eq:delta_theta}
\delta\theta = -\delta \bm{R} \cdot \nabla \theta_{\rm d}.
\end{equation}
Replacing this into \eqref{eq:delta_F1}, using Eqs. (\ref{eq:theta_decomposition}) and (\ref{eq:theta_defect}) and taking into account that $\nabla^{2}\theta_{\rm d}=0$, yields, up to terms of order of the defect core radius $a$:
\begin{multline}\label{eq:delta_F2}
\delta \mathscr{F} 
= \delta \bm{R}\cdot \Big\{ \int {\rm d}A\,\nabla\theta_{\rm d}\left[K\nabla^{2}\theta_{\rm ext}-U'(\theta)\right] \\
+ K \oint {\rm d}s\,(\bm{N}\nabla\theta_{\rm d}-\nabla\theta_{d}\bm{N})\cdot\nabla\theta_{\rm ext} \Big\} + \mathcal{O}(a),
\end{multline}
where we have approximated $|\nabla\theta_{\rm ext}|^{2}\approx 0$. The $\mathcal{O}(a)$ contributions result from the contour integral of the potential energy $U(\theta)$ and can be ignored for sufficiently small defects core. Next, assuming the defect core to be circular and setting up a local system of polar coordinate $(\varrho,\varphi)$ originating at the defect location, so that $\bm{\hat{\varrho}} = -\bm{N}$, $\nabla\theta_{\rm d}=k/a\,\bm{\hat{\varphi}}$ and ${\rm d}s=a\,{\rm d}\varphi$, the contour integral in \eqref{eq:delta_F2} can be straightforwardly calculated:
\begin{multline} 
\oint {\rm d}s\,(\bm{N}\nabla\theta_{\rm d}-\nabla\theta_{d}\bm{N})\cdot\nabla\theta_{\rm ext} \\
= k \int_{0}^{2\pi}{\rm d}\varphi\,(\bm{\hat{\varphi}}\,\bm{\hat{\varrho}}-\bm{\hat{\varrho}}\,\bm{\hat{\varphi}})\cdot\nabla\theta_{\rm ext}
= 2\pi k \nabla_{\perp}\theta_{\rm ext},
\end{multline}
where $\nabla_{\perp}=(-\partial_{y},\partial_{x})$ and we have taken  and assumed $\nabla\theta_{\rm ext}$ so be constant along the core boundary. Furthermore, using the fact that $\nabla\theta_{\rm d}=-\nabla_{\bm{R}}\theta_{\rm d}$, where $\nabla_{\bm{R}}$ represents a gradient with respect to the coordinates of the core, and that $\nabla_{\bm{R}}\theta_{\rm ext}=0$, \eqref{eq:delta_F2} can be rearranged in the form:
\begin{multline}\label{eq:delta_F3}
\delta \mathscr{F} 
= \delta\bm{R}\cdot\Big\{ 2\pi k K \nabla_{\perp}\theta_{\rm ext} \\ 
+ K \int {\rm d}A\,\nabla\theta_{\rm d}\nabla^{2}\theta_{\rm ext} 
+ \nabla_{\bm{R}} \int {\rm d}A\,U(\theta) \Big\}+\mathcal{O}(a).
\end{multline}
Now, taking $\delta\bm{R}=(\dot{\bm{R}}-\bm{v})dt+\mathcal{O}(dt^{2})$ in \eqref{eq:delta_theta}, with $\dot{\bm{R}}=\partial_{t}\bm{R}$ and $\bm{v}$ the flow velocity, we can express the time derivative $\partial_{t}\theta$ as a function of the defect velocity. Namely:
\begin{equation}
\partial_{t}\theta = -(\dot{\bm{R}}-\bm{v})\,\cdot\nabla\theta_{\rm d}.	
\end{equation}
This allows to express the left-hand side of \eqref{eq:integral_eom} in the form:
\begin{equation}
\int {\rm d}A\,\delta\theta\,\partial_{t}\theta =\delta\bm{R}\cdot\left(\int {\rm d}A\,\nabla\theta_{\rm d}\nabla\theta_{\rm d}\right)\cdot(\dot{\bm{R}}-\bm{v})\;.  	
\end{equation}
Finally, combining this with \eqref{eq:delta_F3}, we obtain an equation of motion for the moving defects:
\begin{multline}\label{eq:eom1}
\bm{\zeta}\cdot(\dot{\bm{R}}-\bm{v})
=-2\pi k K\nabla_{\perp}\theta_{\rm ext} \\
- K \int {\rm d}A\,\nabla\theta_{\rm d}\nabla^{2}\theta_{\rm ext} 
-\nabla_{\bm{R}}\int {\rm d}A\,U(\theta).
\end{multline}
where:
\begin{equation}
\bm{\zeta} = \gamma \int {\rm d}A\,\nabla\theta_{\rm d}\nabla\theta_{\rm d}, 
\end{equation}
is an effective drag tensor. As shown in Ref. \cite{Denniston:1996}, this can be explicitly calculated by expressing \eqref{eq:relaxation} in the frame of the moving defects. This yields: $\zeta_{ij} = \zeta \delta_{ij}$, with: 
\begin{equation}
\zeta \approx \pi \gamma k^{2} \log \left(\frac{3.6}{\mathcal{E}}\right),
\end{equation}
with $\mathcal{E}=\gamma a |\dot{\bm{R}}| / K$. In first approximation, $\log (3.6/\mathcal{E})\approx 1$ as a defect typically moves by a few core radii within the nematic relaxational time scale, thus $|\dot{\bm{R}}|\approx a/\tau_{a}$, with $\tau_{a}=\gamma a^{2}/K$.

\begin{figure*}[t]
\centering
\includegraphics[width=0.75\textwidth]{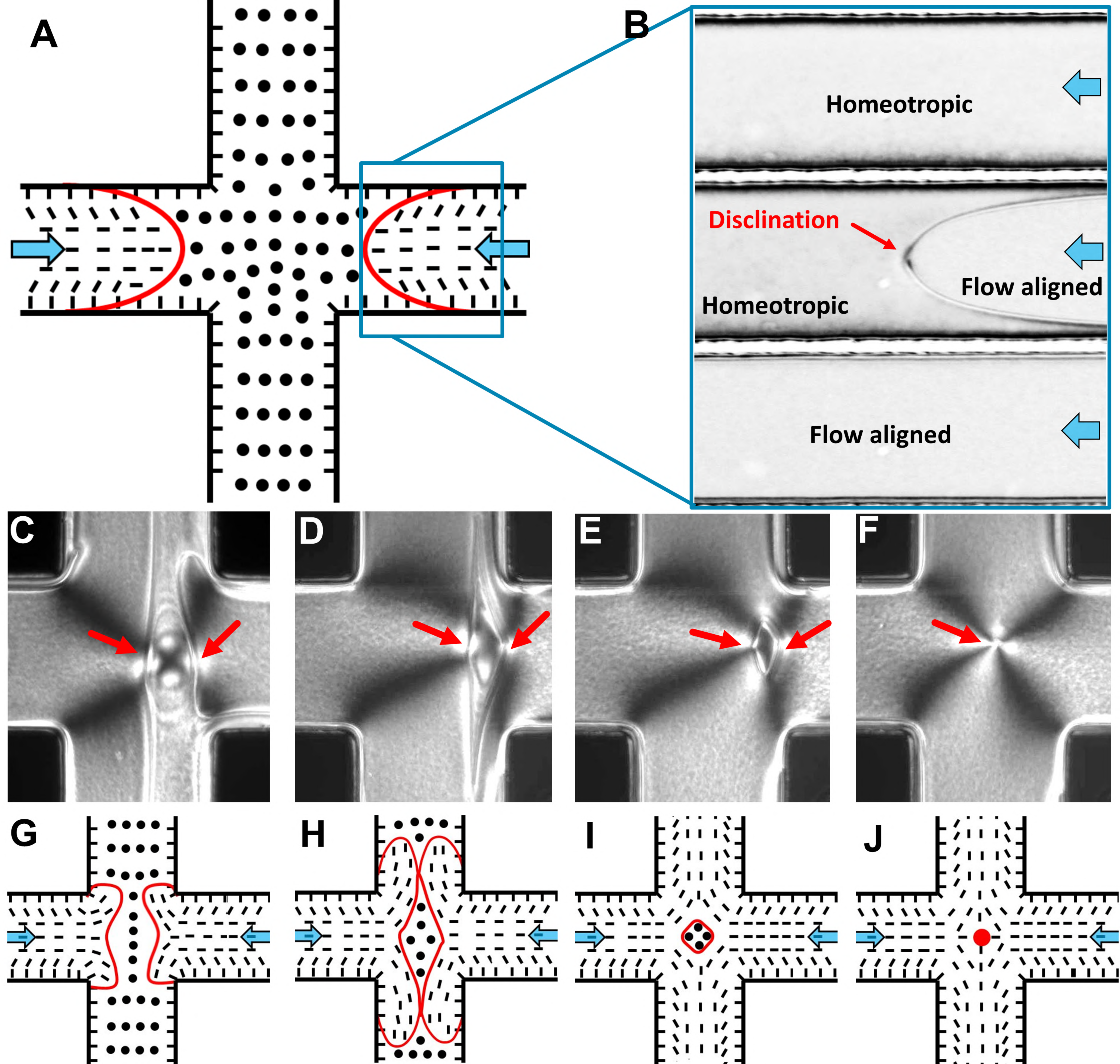}
\caption{\label{fig:nucleation} Dynamics of the defect nucleation at the junction center. (A) Schematic representation of the nematic director field in the in-flow arms of a 4-armed microfluidic junction.  Upon starting the flow, a disclination line connected to the channel walls emerges (shown in red). This disclination separates two distinct director domains – the flow-induced planar orientation (denoted by broken lines) from the surface-induced homeotropic orientation (denoted by the dots). (B) Image sequence showing gradual re-alignment of a homeotropic orientation (top panel) to a flow-aligned director field (bottom panel), imaged using white light. The intermediate snapshots (middle panel) shows the disclination traveling toward the center of the junction. (C-J) Two such disclinations, one from each in-flow arm meet at the junction to form the central defect loop. Polarization optical micrographs (C-F) and schematic representations (G-J) capture the emergence of the central defect loop over time. The broken lines denote flow-aligned director field and the dots show homeotropic anchoring (into the plane of the image). The two disclination lines (indicated by the red arrows) meet at the center. Upon meeting, the singular defects merge into a -1 defect loop, enclosing a homeotropic domain [panels (E,I)]. The defect loop gradually shrinks in size, finally leaving the defect monopole at the junction center [panels (F,J)].}
\end{figure*}

The dynamics of an isolated defect is then dictated by two driving forces: the elastic force, proportional to the elastic constant $K$, which tends to reorient the defect velocity depending to the far field orientation $\theta_{\rm ext}$, and the force $-\nabla_{\bm{R}}\int {\rm d}A\,U(\theta)$, which drives the defect toward the minima of the potential energy. A special scenario, is obtained when $\theta_{\rm ext}$ consists of the orientation field of other topological defects. In this case, one can approximate:
\begin{equation}
\theta(\bm{r}) = \sum_{i}\theta_{\rm d}(\bm{r},\bm{R}_{i}),	
\end{equation} 
where the sum runs over all the defects in the system. Thus, for each of them, $\theta_{{\rm ext},i}(\bm{r})=\sum_{j\ne i}\theta_{d}(\bm{r},\bm{R}_{j})$ and \eqref{eq:eom1} yields Eq. (2), with $\mu_{i}=1/\zeta_{i} \approx 1/(\gamma k_{i}^{2})$. The force $\bm{F}_{i}$ is given by:
\begin{equation}\label{eq:force_flow}
\bm{F}_{i}=-\nabla_{\bm{R}_{i}}\int {\rm d}A\,U(\theta) = - k_{i} \int {\rm d}A\,\frac{\bm{\hat{z}}\times (\bm{r}-\bm{R}_{i})}{|\bm{r}-\bm{R}_{i}|^{2}}\,U'(\theta).  
\end{equation}
Finally, expressing $U'(\theta)$ as given by \eqref{eq:partial_U} we obtain Eq. (3).

As an example of the effect of a high$-\mathrm{Er}$ flow on the motion of a defect, we consider the simple case of a $4-$arm junction, whose velocity field is approximated by Eqs. (\ref{eq:cross_flow_v}). The corresponding strain-rates and vorticity are given by:
\begin{subequations}\label{eq:cross_flow_strain_rate}
\begin{align}
u_{xx} &= \frac{8v_{0}}{3 w^{3}}\,\left[w^{2}-3(x^{2}+y^{2})\right], \\[5pt]
u_{xy} &= 0,\\[5pt]
\omega_{xy} &= \frac{16 v_{0}}{w^{3}}\,x y.
\end{align}
\end{subequations}
Fig. \ref{fig:defect_force_field} shows the force field experienced by a $\pm 1$ disclination at the center of a $4-$arm junction and calculated via Eqs. (\ref{eq:partial_U}), (\ref{eq:force_flow}) and (\ref{eq:cross_flow_strain_rate}). As consequence of such a force field, negative defects are attracted by the central stagnation point, while positive disclinations are repelled toward the channels.

\section{\label{sec:new}Dynamics of the Defect Nucleation at the Junction Center}

Homeotropic microfluidic channels, like the ones used in the present work, support multiple nematic configurations, either stable or metastable, in absence of any flow. Depending on the deformation of the director close to the channel corners, these different possible configurations correspond to different free energy values. Specifically, the channel aspect ratio (channel width/height) and the curvature (sharpness) at the corners determine whether nematic defects will be real bulk or virtual \cite{Sengupta:2013,Sengupta:2014}. Irrespective of these multiple initial conditions, when we initiate flow in the microchannel, a pseudo-planar structure first emerges, and then stabilizes into a flow-aligned director configuration. Specifically, we observe three different flow regimes within microchannels having rectangular cross-section depending upon the Ericksen numbers \cite{Sengupta:2013}. The flow regime relevant to the current work is the “high” Ericksen number regime, where the nematic profile evolves into a flow-aligned state, with the director oriented primarily along the channel length. The alignment initiates close the channel inlet and propagates downstream as the nematic director gets distorted. Further downstream, the director field remains relatively undisturbed. Thus, a linear microchannel develops two director domains: upstream, a flow-aligned director domain; and downstream, an intact homeotropic domain (Fig. \ref{fig:nucleation}A,B). The two director field domains are separated by a disclination which spans the width of the channel and connects at the surfaces of the channel walls. Two disclinations, one in each of the in-flow arms, travel downstream (Fig. \ref{fig:nucleation}B) and meet at the central junction region. 

The disclinations recombine at the junction to form the central defect loop. As presented in the polarization optical micrographs Fig. \ref{fig:nucleation}(C-F) and the corresponding director field schematics Fig. \ref{fig:nucleation}(G-J), upon meeting, they merge into a single $-1$ defect loop of winding number $-1/2$. Owing to the singularity in the flow field close to the center of the junction, the surface-induced homeotropic texture persists where the opposite streamlines intersect(flow speed $\to$ 0). Consequently, the $-1$ defect loop encloses a homeotropic domain at its center, with flow-aligend director field surrounding it (panels (E) and (I) in Fig. \ref{fig:nucleation}). Thereafter, the defect loop shrinks in size to minimize the free energy, till it reaches the final morphology of a monopole at the junction center. This is presented in Fig. \ref{fig:nucleation}(F,J). The mechanism described here can however differ with the curvature and the geometry of the microchannel cross-section. In absence of flow, strong geometrical curvatures can support peripheral disclinations (running parallel to the channel walls) \cite{Sengupta:2014}, which under specific conditions, separate the surface-induced homeotropic texture from stable pseudo-planar textures \cite{Pieranski:2016a,Pieranski:2016}. As the nematic flow is initiated, the homeotropic texture can be eliminated in favor of the pseudo-planar texture due to the movement of the peripheral defect line orthogonal to the flow direction. We believe, in such a setting, the emergence of the central defect loop will be additionally influenced by the peripheral disclinations \cite{Pieranski:2016} and by potential generation of umbilics \cite{Pieranski:2014}.

\begin{figure}[t]
\centering
\includegraphics[width=\columnwidth]{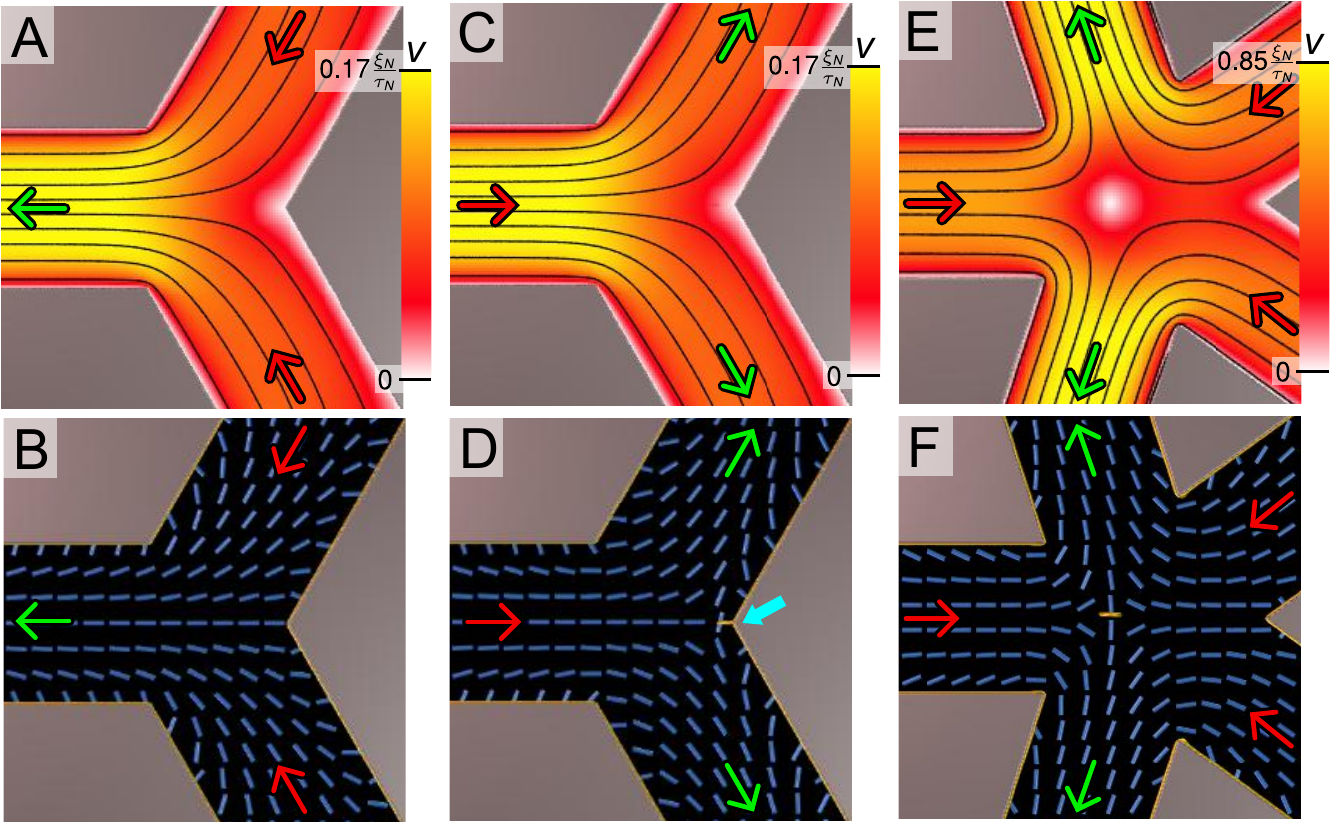}
\caption{\label{fig:35junction} Simulations of (A-D) 3 junctions and
(E,F) 5 junctions of nematic microchannels. Panels in upper row show the flow
profile with the color bar representing the speed in a junction, and
streamlines representing the velocity direction. Corresponding director field (blue
rods) and scalar order parameter field (isosurface at $S=0.4$) are shown in the low row panels. 
In a 3-junction we show (A,B) regime of 2 inlet
flows and (C,D) regime of two outlet flows. With two inlet flows in a
3-junctions there are no nematic defects present. However, if there is
only one inlet flow, the nematic forms a $-\frac{1}{2}$ defect line that
is pinned to the corner of a junction (marked by a blue arrow in D).
(E,F) 5-junction with 3 inlet flows. The center of the junction
resembles a 4-junction and shows a small defect loop coinciding with the
stagnation point.}
\end{figure}

\section{\label{sec:6}Numerical Simulations of Odd$-$arm Junction}

%In the main text, we show nematic configurations in junctions of 4, 6, and 8 microchannels. In Fig.~\ref{fig:35junction} we show numerical simulations of junctions of odd number of nematic microchannels. In a $3-$arm junction, the stagnation point occurs at the corner of the junction. In the regime of 2 outlet flows, a nematic $-1/2$ pined defect line occurs at same corner. In a $5-$junction, there are 2 stagnation points: one close to the center of the junction and one pinned to the corner like in a $3-$arm junction. In Fig.~\ref{fig:35junction} we show a $5-$arm junction with 3 inlet flows and 2 outlet flows. The nematic configuration in such junction resembles the nematic configuration in a $4-$arm junction with a small defect loop/point defect located at the position of a stagnation point with a $2-$fold rotational symmetry.

In the main text, we show nematic configurations in junctions of 4, 6,
and 8 microchannels. In Fig.~\ref{fig:35junction} we show numerical
simulations of junctions of odd number of nematic microchannels. In a
$3-$arm junction, the stagnation point occurs at the corner of the
junction. The emergence of a defect at the stagnation point is
conditioned by the inflow/outflow regime. In the regime of 2 outlet
flows, a nematic $-1/2$ pined defect line occurs at same corner, while
in the regime of 2 inlet flow there is no nematic defect at the
stagnation point.

In a $5-$junction, there are 2 stagnation points: one close to the
center of the junction and one pinned to the corner like in a $3-$arm
junction. In Fig.~\ref{fig:35junction} we show a $5-$arm junction with 3
inlet flows and 2 outlet flows. The nematic configuration in such
junction resembles the nematic configuration in a $4-$arm junction with
a small defect loop of topological charge $-1$ overlaping with the
stagnation point which has the same topology as in a $4-$junction.

\end{document}